      [1999/12/01 v1.4c Il Nuovo Cimento]
\documentclass[varenna]{cimento}

\usepackage{amsmath}
\usepackage{amssymb}
\usepackage{amsfonts}
\usepackage{amsbsy}

\usepackage{graphicx}  
\usepackage{afterpage}
\usepackage[utf8]{inputenc}

\def\ket#1{{\ensuremath{\left|#1\right\rangle}}}

\def\ketbra#1#2{{\ensuremath{\left|{#1}\left>\vphantom{#1}%
  \vphantom{#2}\right<{#2}\right|}}}

\def\braopket#1#2#3{{\ensuremath{\left<{#1}\left|\vphantom{#1}%
  #2\vphantom{#3}\right|{#3}\right>}}}

\def\abs#1{\ensuremath{\left|#1\right|}}

\def\vect#1{{\ensuremath{\overset{\rightarrow}{#1}}}}

\def\dadt{\ensuremath{\mathrm{\frac d{dt}}}}

\def\unite#1{\ensuremath{~\mathrm{#1}}}

\def\cf{cf.\ }

\author{P.~Clad\'e}

\institute{Laboratoire Kastler Brossel, Universit\'e Pierre et Marie Curie, ENS, CNRS; 4, place Jussieu; 75005 Paris}

\title{Bloch oscillations in atom interferometry}

\begin{document}

\maketitle

\begin{abstract}
In Paris, we are using an atom interferometer to precisely measure the recoil velocity of an atom that 
absorbs a photon. In order to reach a high sensitivity, many recoils are transferred to atoms using the Bloch oscillations technique. 
In this lecture, I will present in details this technique and its application to high precision measurement. I will especially describe in details
how this method allows us to perform an atom recoil measurement at the level of $1.3 \times 10^{-9}$. This measurement is used in the most precise determination of the fine structure constant that is independent of quantum electrodynamics. 
\end{abstract}

\section{Introduction}

Atom interferometry \cite{Borde1989} is now used in many ways to perform precise measurements \cite{Cronin2009}. The most advanced application of atom interferometry is gravity measurements. A sensitivity of $1.4\times10^{-8}g$ at 1~s is obtained \cite{LeGouet:2007} with an accuracy at the level of $10^{-9}$ \cite{Peters1999}. Higher sensitivities in differential measurements are obtained \cite{McGuirk2002} and are used to perform measurements of the Newtonian Constant of Gravity \cite{fixler2007, Lamporesi2008}.

Atom interferometers are also used to precisely measure the atom recoil velocity in order to get a value of the fine structure constant. This idea dates back to the 1990s. The group of S.~Chu at Stanford was a pioneer in this field \cite{Weiss92}. In 2002 a value of the Cs recoil velocity with an uncertainty of $14\times 10^{-9}$ was published\cite{Wicht:02}. This experiment is now under the supervision of H.~Müller. They are now able to reach an accuracy of $4\times 10^{-9}$\cite{Lan2013}. In 2011, we published, at the Laboratoire Kastler Brossel, a value with an uncertainty of about $1.3\times 10^{-9}$\cite{Bouchendira2011}. The principle of this experiment consists in using an atom interferometer to measure the change of velocity of the atoms after an acceleration of many recoils. Those recoils are transferred using the method of Bloch oscillations \cite{BenDahan} which is very efficient. The number $N$ of recoils that are transferred is indeed an important parameter because the resolution of the interferometer is proportional to $N$. 

\medskip

The two lectures given at this Enrico Fermi school will focus specifically on the Bloch oscillation method, which is a tool that we have studied and used in the group of F.~Biraben at Laboratoire Kastler Brossel since 1998. In the first lecture, I will describe how, from basic principles of atom light interaction, we can understand and modelize the phenomenon of Bloch oscillations. This will be done using both the particle and the wave points of view. In the second lecture, I will present how Bloch oscillations are used in high precision atom interferometers. Three different examples will be used : the first one is on the measurement of gravity, the second one on the use of Bloch oscillations to enhance atomic beamsplitters and the third example will be dedicated to the state-of-the-art measurement of the recoil velocity and its application to fundamental physics.






\begin{center}
\Large\textbf Lecture I : Principle of Bloch oscillations
\end{center}


Before describing the physics behind Bloch oscillation, I want to give key ideas of the experiment carried out in Paris. 
The aim of the experiment is the precise measurement of the recoil velocity of an atom that absorbs a photon. A simple scheme for this experiment would be to prepare atoms with a well known velocity, then transfer to those atoms one photon recoil and then measure the change of velocity. Two approaches are used in order to reach a high sensitivity : the first one is to use an atom interferometer to precisely measure the velocity of atoms. The second one consists in transferring many recoils to the atoms using the Bloch oscillations technique that we will describe in details. 

If we denote $N$ the number of recoils that we are able to transfer to the atoms and $\sigma_v$ the sensitivity in velocity of the atom interferometer, then the final sensitivity on the recoil velocity will be 
\begin{equation}
\label{eq:sensitivity}
\sigma_{v_r} = \frac{\sigma_v}N
\end{equation}

Let us give some numbers for the rubidium atom : the recoil velocity is about $6~\unite{mm/s}$ for a photon at $780\unite{nm}$. Using the same wavelength, the Doppler effect associated with this velocity is $7.5\unite{kHz}$. 

The basic components of this measurement are two-photon counter-propagating Raman transitions. Such transitions are used both for the velocity sensor and the recoil transfer (Bloch oscillations). 

\section{Two photon Raman transition : recoil and Doppler effect}

The atom interferometer relies on the measurement of the Doppler effect on an atomic transition. The acceleration relies on the momentum transfer from light to atoms. In both cases we use a two-photon Raman transition (lambda scheme). It is mandatory to use such a transition, because a single-photon transition will induce spontaneous emission. This spontaneous emission will induce a large broadening of the transition (on the order of $10\unite{MHz}$ for rubidium atoms). It will also randomly change the velocity of the atoms.

\subsection{Principle of Raman transitions}

The principle of the Raman transition consists in shining the atoms with two lasers of different frequencies. Atoms will then be able, in a single process, to absorb a photon from one laser and emit a photon into the mode of the second laser. The resonance applies only to the initial and final state, the "intermediate" state where the atom has absorbed one photon being out of resonance. Because the photon is simultaneously emitted, the spontaneous emission is strongly reduced. 

\begin{figure}
\begin{center}
\includegraphics[width=.7\linewidth]{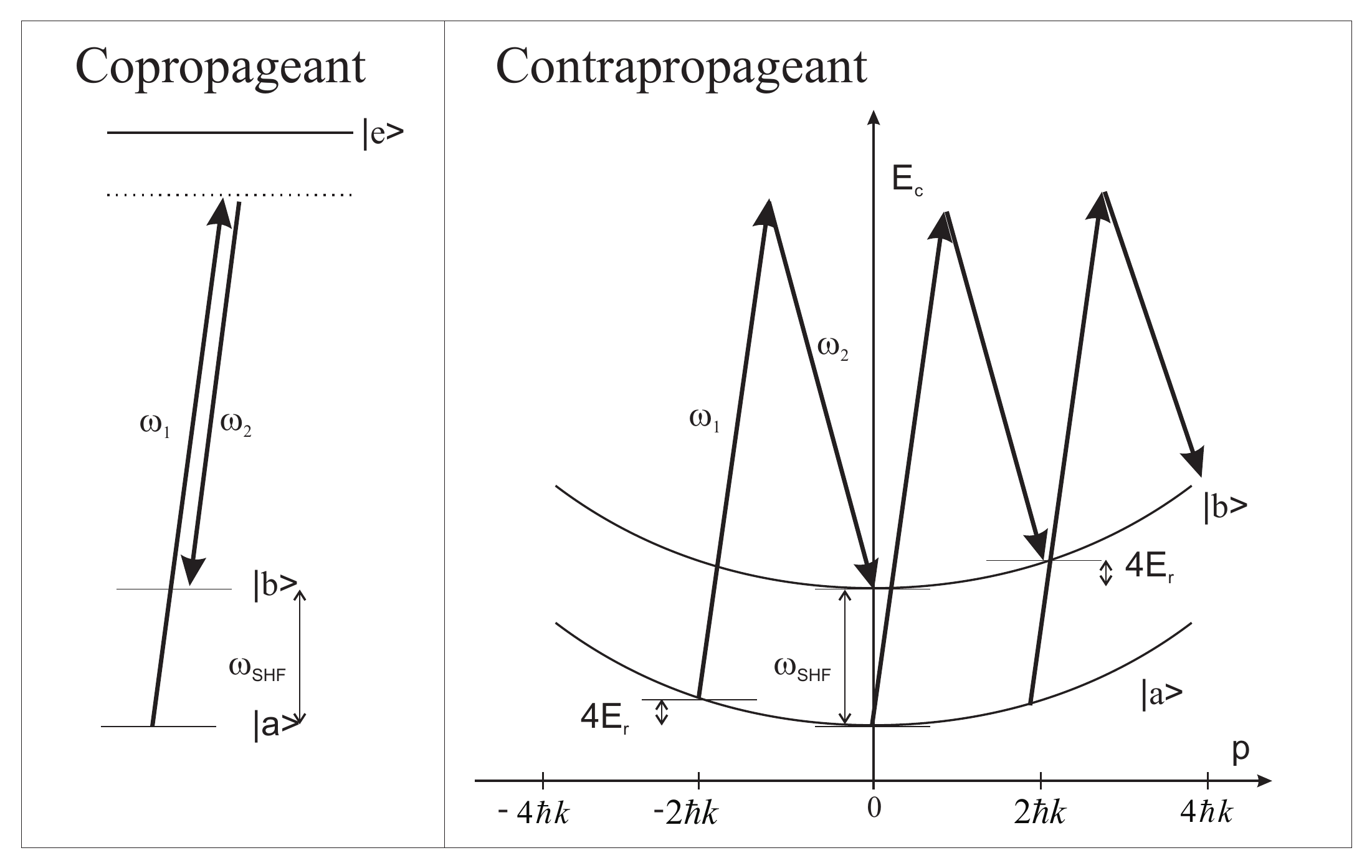}
\end{center}
\caption{\label{fig:Raman}Raman transition in two different configuration : the co-propagating (left) and the counterpropagating (right).}
\end{figure}

We will describe the Raman transition in a simple three-level scheme depicted in Fig.~\ref{fig:Raman}. The resonance condition results from the total conservation of the energy and momentum. Let us call $v_i$ the initial velocity of the atom. This condition is then : 
\begin{eqnarray}
E_a + \frac12mv_i^2 + \hbar\omega_1 &=& E_B + \frac12mv_f^2 + \hbar\omega_2 \\
m\vect{v_i} +\hbar \vect{k_1} &=& m\vect{v_f} +\hbar \vect{k_2}
\end{eqnarray}
where $v_f$ is the final velocity, $\omega_{1,2}$ the pulsation of the lasers and $\vect{k}_{1,2}$ their wave-vector. Let us call $\delta = \omega_1 - \omega2 - (E_a - E_b)/\hbar$ the detuning of the Raman transition. The resonance condition then gives a value of $\delta$ : 
\begin{equation}
\delta = \left(\vect{k_1} - \vect{k_2}\right)\cdot\left(\vect{v_i} + \frac\hbar{2m}\left(\vect{k_1} - \vect{k_2}\right) \right)
\end{equation}

This equation contains two terms : the Doppler shift and the recoil effect. As we can see on Fig.~\ref{fig:Raman}, in the counterpropagating scheme ($\vect{k_1} \simeq - \vect{k_2}$), in order to be resonance with the atoms at rest, one needs to shift the laser by $4E_r/\hbar$, where $E_r$ is the recoil energy defined by:
\begin{equation}
E_r = \frac{\hbar^2k^2}{2m}
\end{equation}

\subsection{Hamiltonian of the Raman transition}
\label{sec:hamilRaman}

The transition probability can be calculated using the dressed atom picture. We have two lasers at pulsations $\omega_1$ and $\omega_2$. The number of photons in each mode is given by $\hat{N_1}$ and $\hat{N_2}$. We will restrict the atom to three states : the two hyperfine states of the ground state $\ket{a}$ and $\ket{b}$ and the excited state $\ket{e}$. The Hamiltonian without interactions $H^{(0)}$ contains three terms : the kinetic energy of the atom, the internal energy of the atom and the photon's energy : 
\begin{equation}
H^{(0)} = \frac{\hat{p}^2}{2m} + E_a \ketbra{a}{a} + E_b \ketbra{b}{b} + E_c \ketbra{c}{c} + \hat{N_1}\hbar\omega_1 + \hat{N_2}\hbar\omega_2
\end{equation}

The eigenstates of this Hamiltonian are \ket{i,\vect{p}, n_1, n_2}, where $i$ is the internal state, $\vect{p}$ the momentum of the atom and $n_l$ the number of photons in the mode $l$. 

\begin{figure}
\begin{center}
\includegraphics[width=.7\linewidth]{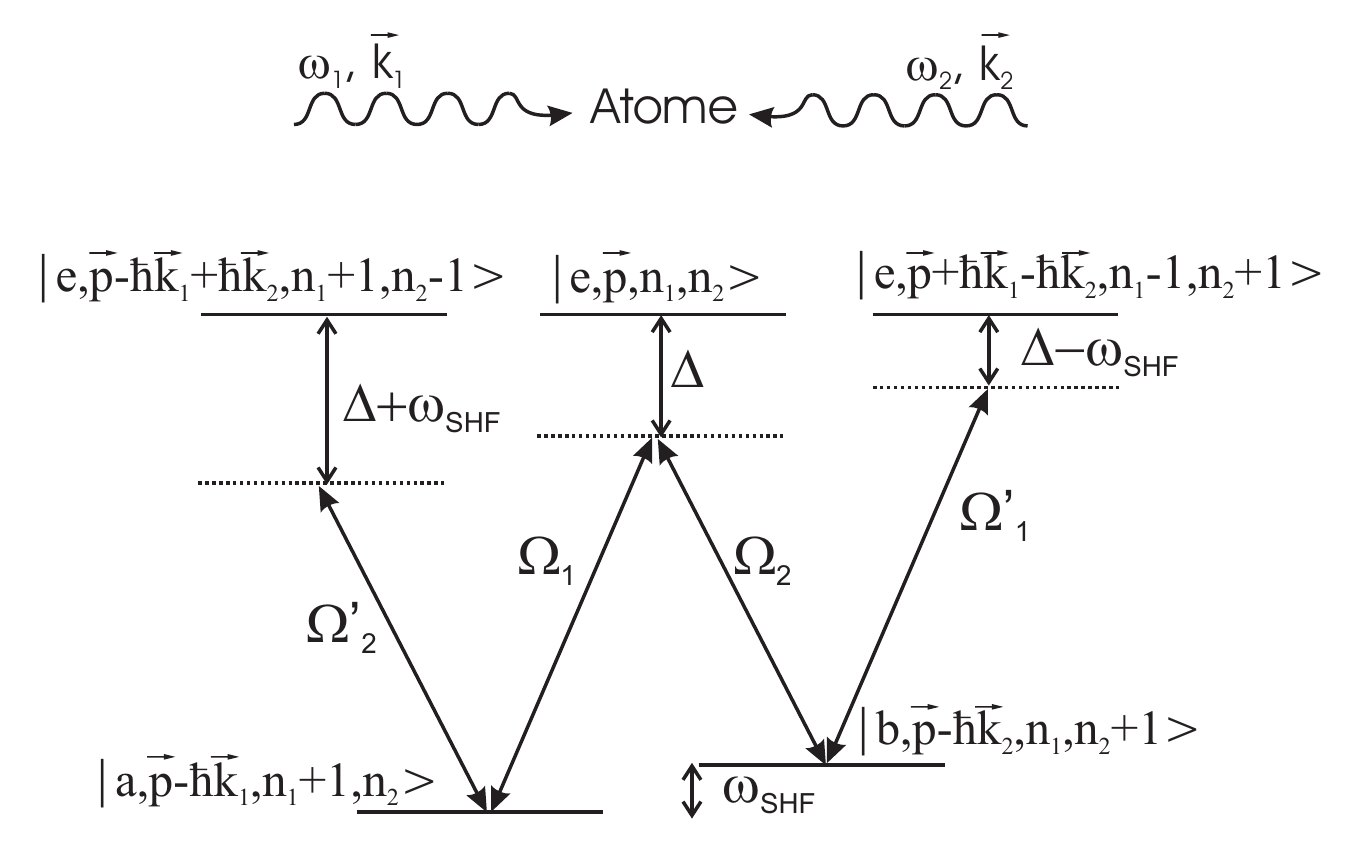}
\end{center}
\caption{\label{fig:RamanHabille}Raman transition in the dressed atom picture}
\end{figure}

Let us denote by $\hat{V} = -\vect{\hat{d}}\cdot \vect{\hat{E}}$ the interaction between atoms and field, where \vect{\hat{d}} is the dipole operator (in our case \vect{\hat{d}} will couple \ket{a} and \ket{b} to \ket{e}), and $\vect{\hat{E}}$ the operator corresponding to the electric field.
The photon field $1$ will couple the state \ket{e,\vect{p}, n_1, n_2} to the state \ket{a,\vect{p}-\hbar\vect{k_1}, n_1+1, n_2}. We can therefore define a first Rabi frequency : 
\begin{equation}
\frac{\Omega_1}2 = \braopket{e,\vect{p}, n_1, n_2}{\hat V}{a,\vect{p}-\hbar\vect{k_1}, n_1+1, n_2}
\end{equation}
Similarly, we can define the Rabi frequencies $\Omega_2$, $\Omega^\prime_1$ and $\Omega^\prime_2$ represented in the Fig.~\ref{fig:RamanHabille}. 

In the following, we will consider only the two states \ket{a,\vect{p}-\hbar\vect{k_1}, n_1+1, n_2} and \ket{b,\vect{p}-\hbar\vect{k_2}, n_1, n_2+1}, that we will call \ket{a} and \ket{b} respectively. In the basis of those two states, the Hamiltonian is simply:
\begin{equation}
H^{(0)} = \hbar\begin{pmatrix}\delta& 0 \\
0 & 0
\end{pmatrix}
\end{equation}
where $\delta = \omega_1 - \omega_2 -\omega_\mathrm{SHF}$. 

In typical applications, the detuning $\delta$ is in the kHz range, the coupling $\Omega$ in the MHz range and the detuning $\Delta$ in the GHz range. We can therefore consider that \textit{i)} the two states \ket{a} and \ket{b} are quasi-degenerate and \textit{ii)} the perturbation theory can be applied. 

If $i$ and $j$ represents two state in the doublet \ket{a} and \ket{b}, then the Hamiltonian at the lowest order in V that couples the two states is\footnote{\cf Landau \& Lifchitz \cite{Landau}, \S38.}
\begin{equation}
H_{ij} = H^{(0)}_{ij} + \sum_{\alpha\in \left\{\mathrm{\mbox{\centering excited states}}\right\}}
\frac{V_{i\alpha}V_{\alpha j}}{H^{(0)}_{ii}-H^{(0)}_{\alpha\alpha}}
    \label{HamiltonienEffectif}
\end{equation}
where the sum over $\alpha$ is on the three excited dressed states.

\def\omegaSHF{\omega_\mathrm{SHF}}
Finaly, in the base of (\ket{a},\ket{b}) the Hamiltonian is :
\begin{equation}
\hbar\begin{pmatrix}
  \delta + \frac{(\vec p - \hbar \vec{k_1})^2}{2m\hbar}  - \frac{|\Omega_1|^2}{4\Delta} - \frac{|\Omega^\prime_2|^2}{4(\Delta+\omegaSHF)}&
  \frac{\Omega_1\Omega_2^*}{4\Delta} \\
  \frac{\Omega_1^*\Omega_2}{4\Delta} &
  \frac{(\vec p - \hbar \vec{k_2})^2}{2m\hbar}  - \frac{|\Omega_2|^2}{4\Delta} - \frac{|\Omega^\prime_1|^2}{4(\Delta-\omegaSHF)} \\
\end{pmatrix}
\label{HamiltonienMatrice}
\end{equation}

Using the following notations
\begin{eqnarray}
    \delta_a &=& \frac{|\Omega_1|^2}{4\Delta} + \frac{|\Omega^\prime_2|^2}{4(\Delta+\omegaSHF)} \\
    \delta_b &=& \frac{|\Omega_2|^2}{4\Delta} + \frac{|\Omega^\prime_1|^2}{4(\Delta-\omegaSHF)} \\
    \Omega^{\mathrm{eff}}_{a\rightarrow b} &=& \frac{\Omega_1\Omega_2^*}{2\Delta} \\
    \hbar\delta_{\mathrm{Doppler}}(\vec{p}) &=& \frac{(\vec p - \hbar \vec{k_1})^2}{2m} -\frac{(\vec p -
\hbar \vec{k_2})^2}{2m} = \hbar\left(\vec{k_2} - \vec{k_1}\right).\frac{\vec{p}}m
\end{eqnarray}
we obtain the following Hamiltonian : 
\def\Omegaeffab{\ensuremath{\left.\Omega^{\mathrm{eff}}_{a\rightarrow b}\right.}}
\begin{equation} \label{Eq:HamiltonienEffectif}
    \hbar
\begin{pmatrix}
  \delta-\delta_a & \Omega^{\mathrm{eff}}_{a\rightarrow b}/2 \\
  \Omega^{\mathrm{eff}*}_{a\rightarrow b}/2 & -\delta_b - \delta_{\mathrm{Doppler}}\\
\end{pmatrix}
\end{equation}

The energy $\hbar\delta_a$ and $\hbar\delta_b$ are the light shift of states \ket{a} and \ket{b}; $\Omega^{\mathrm{eff}}_{a\rightarrow b}$ is the effective coupling between \ket{a} et \ket{b}.

In this section, we have presented the simple case of a three-level system. This model can be generalized to any atoms with hyperfine structure: in equation \ref{HamiltonienEffectif}, the sum is on all the hyperfine states whose matrix elements are calculated using Clebsch-Gordan coefficients. An effective Hamiltonian between all the different hyperfine states of the ground state can then be calculated. It will depend on the polarisation of the Raman laser beams. This precise model is also useful for the calculation of systematic effects coming from the light shift (see section \ref{sec:effet_sys}).

\subsection{Raman transition in the same internal state}
In the previous paragraph, we have described the case of a Raman transition between two different hyperfine states. By absorbing a photon from laser $1$ and emitting a photon into the mode of laser $2$, the atoms jump from \ket{a} to \ket{b} and are accelerated by two photon recoils. If we want to further accelerate atoms, using a similar process, we need to repeat this process by inverting the direction of laser $1$ and $2$. This is the method used in the group of S.~Chu for their first measurement of the ratio $h/m$ \cite{Weiss94}. 

Another method for accelerating atoms consists in using a similar Raman transition, but keeping the same internal state. In this case the two lasers' pulsations $\omega_1$ and $\omega_2$ will be almost equal ($\abs{\omega_1 - \omega_2} \ll \Omega$). The only difference between the initial and final state will be the kinetic energy. The state \ket{a, \vect{p}, N_1, N_2} is then coupled to the state \ket{a, \vect{p} + 2\hbar\vect{k}, N_1-1, N_2+1} and \ket{a, \vect{p} - 2\hbar\vect{k}, N_1+1, N_2-1}, with an effective coupling rate $\Omega^{\mathrm{eff}}$ given by :
\begin{equation}
\Omega^{\mathrm{eff}} = \frac{\Omega^2}{2\Delta}
\end{equation}

We should however note that the state \ket{a, \vect{p} - 2\hbar\vect{k}, N1+1, N2-1} is also coupled to the state \ket{a, \vect{p} - 4\hbar\vect{k}, N1+2, N2-2} and so on. Compared to the Raman transition where only two states where coupled together, we have to consider an infinite number of states labeled with an integer number $l$:
\begin{equation}
\ket{\vect{p}, l} = \ket{a, \vect{p} - 2l\hbar\vect{k}, N1+l, N2-l}
\end{equation}

If the initial state has a well known momentum $\vect{p}$, then it will remain in the basis of the \ket{\vect{p}, l} states. The Hamiltonian can then be reduced to a matrix. 
\begin{equation}
H = \begin{bmatrix}
    \ddots &    \kappa &    0       &    \ldots       & 0    & \ldots& 0 \\
    \kappa &  q_{-l}^2 & \kappa     & \ddots        &    \vdots   &    & \vdots     \\
    0       &   \kappa &   \ddots &    \kappa  &  0&  \ldots &   0 \\
     \vdots   &   \ddots  &   \kappa  &   q_0^2 &   \kappa  &   \ddots  & \vdots\\
     0 &   \ldots    &       0   &   \kappa& \ddots&\kappa& 0  \\
     \vdots       &       &    \vdots       & \ddots  &\kappa & q_l^2 & \kappa \\
    0       &    \ldots   &   0 &      \ldots     &    0  & \kappa & \ddots \\
\end{bmatrix}\label{eq:supermatrice}
\end{equation}
where $q_l = \frac{\left(p+2l\hbar k\right)^2}{2m} + l\hbar\left(\omega_1 - \omega_2\right) + \Omega^\mathrm{eff}$ and $\kappa = \frac{\hbar\Omega^\mathrm{eff}}2$. This Hamiltonian is usually written using dimensionless units, where $\hbar=1$ and the energy scale is given by the recoil energy $E_r = \hbar^2k^2/2m$. 

\section{Atom in an optical lattice}

\subsection{Hamiltonian of an atom in an optical lattice}
In section \ref{sec:hamilRaman}, we have written the one-photon coupling in the basis of internal and momentum eigenstates. The effective coupling of the  Hamiltonian of the previous section is written in the same basis. We can transform it back and write a coupling operator. This effective operator will contain two terms, one corresponding to the transition $\ket{p} \rightarrow \ket{p + 2\hbar k}$, and the other, to the transition $\ket{p} \rightarrow \ket{p - 2\hbar k}$.
\begin{equation}
\hat{V}^\mathrm{eff} = \hbar\frac{\Omega^\mathrm{eff}}2 \left(e^{2ik\hat x} + e^{-2ik\hat x}\right) = \hbar\frac{U_0}4 \left(e^{2ik\hat x} + e^{-2ik\hat x}\right)
\end{equation}
This coupling is written in the semi-classical limit of a high number of photons, where we have removed the quantized electro-magnetic field. The energy $U_0$ is the peak-to-peak amplitude of the coupling.

The full Hamiltonian, in the case where $\omega_1 = \omega_2$, can therefore be written as: 
\begin{equation}
\hat{H} = \frac{p^2}{2m} + \hbar \frac{U_0}2\left(1 + cos(2k\hat x)\right)
\label{eq:hamil_periodic}
\end{equation}

This Hamiltonian can be directly understood as the Hamiltonian of a particle in the periodic potential created by an optical lattice. In this semi-classical picture, the light shift of a single laser of intensity $I$ will displace the level by an energy equal to $\hbar  \Omega^\mathrm{eff} /2$, therefore the lattice, with an intensity $I(x) = 2I \left(1 + cos(2k x)\right)$, will induce a periodic potential given by eq.~\ref{eq:hamil_periodic}. 

\subsection{The Bloch theorem}
The aim of this part is to find the eigenstates of the Hamiltonian of eq.~\ref{eq:hamil_periodic}. The problem of the eigenstates of a periodic Hamiltonian is a well-known problem that was solved in solid-state physics in the late 1920s by Bloch \cite{Bloch}. 

This Hamiltonian is invariant under a translation of a multiple of the distance $d=\lambda /2$. In other words, it means that it commutes with the translation by $d$ operator: 
\begin{equation}
\hat{T}_d = e^{i\frac{\hat p d}\hbar}
\end{equation}

The consequence is that it is possible to find a base of eigenstates of $H$ that are eigenstates of $\hat{T}_d$. The eigenvalues of $\hat{T}_d$ are all complex numbers with unitary modulus. The eigenvalues can be written as $e^{i q d}$, where $q \in ]-\frac\pi d, \frac\pi d] $. The corresponding eigenstates are
\begin{equation}\label{eq:FonctionBloch}
    \psi_q(x) = u_q(x) e^{iqx}
\end{equation} where $u(x)$ is a periodic function. This result is known as the Bloch theorem.

It is hard to find a correct physical interpretation of the number $q$ that we have introduced. One can do an analogy with the case of a free system. In this case, the system is invariant by any translation. The conserved quantity is the momentum. When the system is quantized, this quantity corresponds to the momentum operator $\hat{p}$. The eigenstates of the free Hamiltonian are plane waves. The momentum correspond then to the gradient of the phase of the plane wave (taking $\hbar=1$). In the case of a periodic Hamiltonian, the momentum is not conserved, it is no longer a good quantum number. The good quantum number is the value $q$ that we have introduced. It is called the quasi-momentum. It is not the phase gradient, but the variation of the phase over the distance $d$ divided by $d$. 

The phase is always defined modulo $2\pi$. This does not lead to a problem when calculating the momentum using the phase gradient. However, in the discrete case used to define the quasi momentum, we have to take into account the fact that the phase is defined modulo $2\pi$. Therefore, the quasi-momentum $q$ is defined modulo $2\pi/ d$. The choice of $q \in ]-\frac\pi d, \frac\pi d] $ is purely conventional. The best representation being indeed the phase of the complex number $e^{i q d}$, i.e. a number of a circle. 

\subsection{Band structure}

Using the Bloch theorem, one can calculate the eigenstates and eigenvalues of the Bloch Hamiltonian. For a given $q$, because of the periodicity of the function $u_q$, the Hamiltonian has a discrete number of eigenstates. Let us call $\ket{n,q}$ those eigenstates, $\psi(n,q)(x)$ the wave functions and $E_n(q)$ the eigen-energies. 

The eigenvalues can be calculated by numerically solving the discrete Hamiltonian of eq.~\ref{eq:supermatrice}. In order to solve this matrix, we need to truncate this matrix. It can be shown that we need to keep values of $l$ smaller than $l_{max}$ with $l_{max} \gg \left(\frac{U_0}{E_r}\right)^{\frac14}$\cite{Peik}.

\begin{figure}
\includegraphics[width=.49\linewidth]{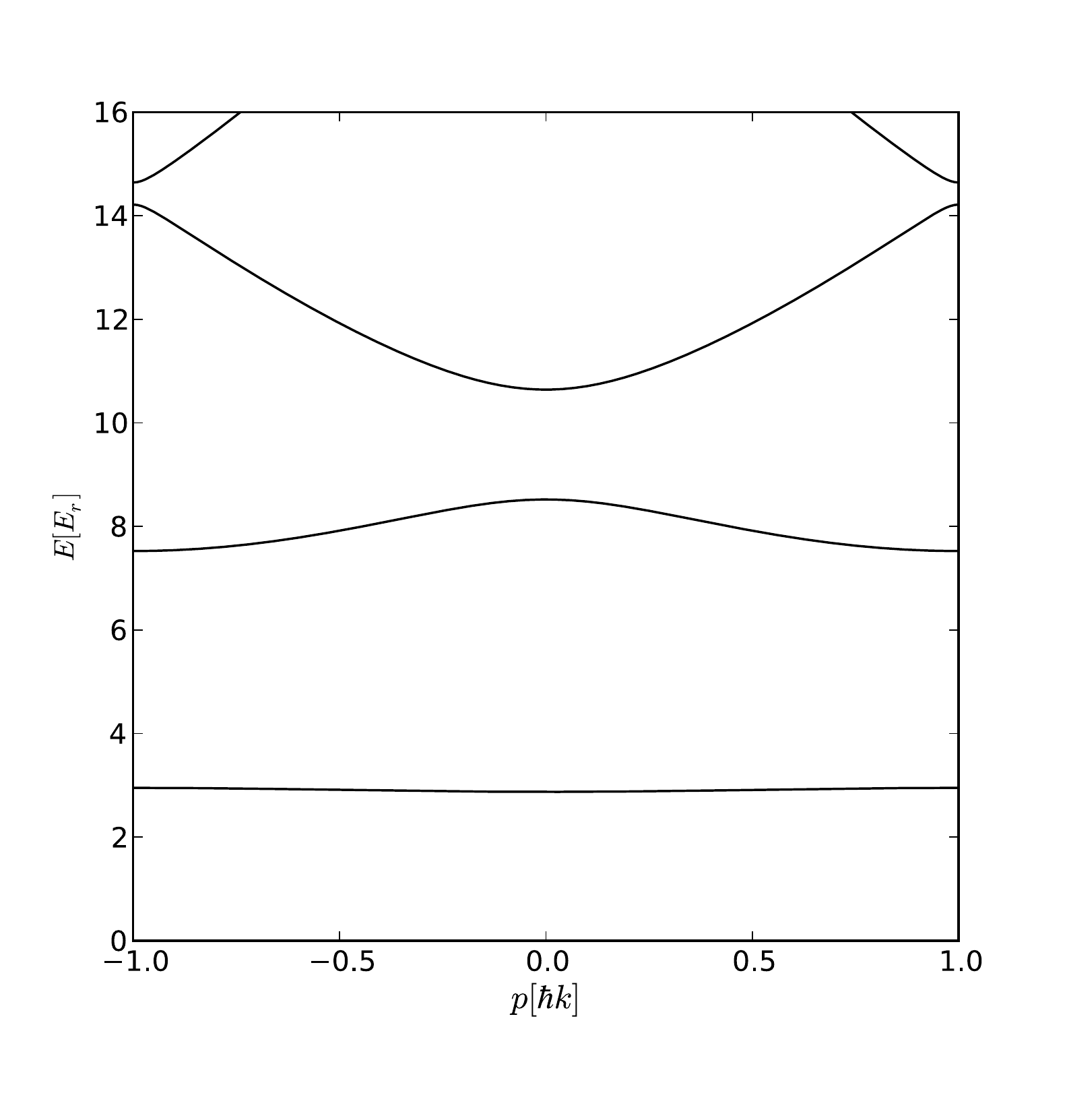}
\includegraphics[width=.49\linewidth]{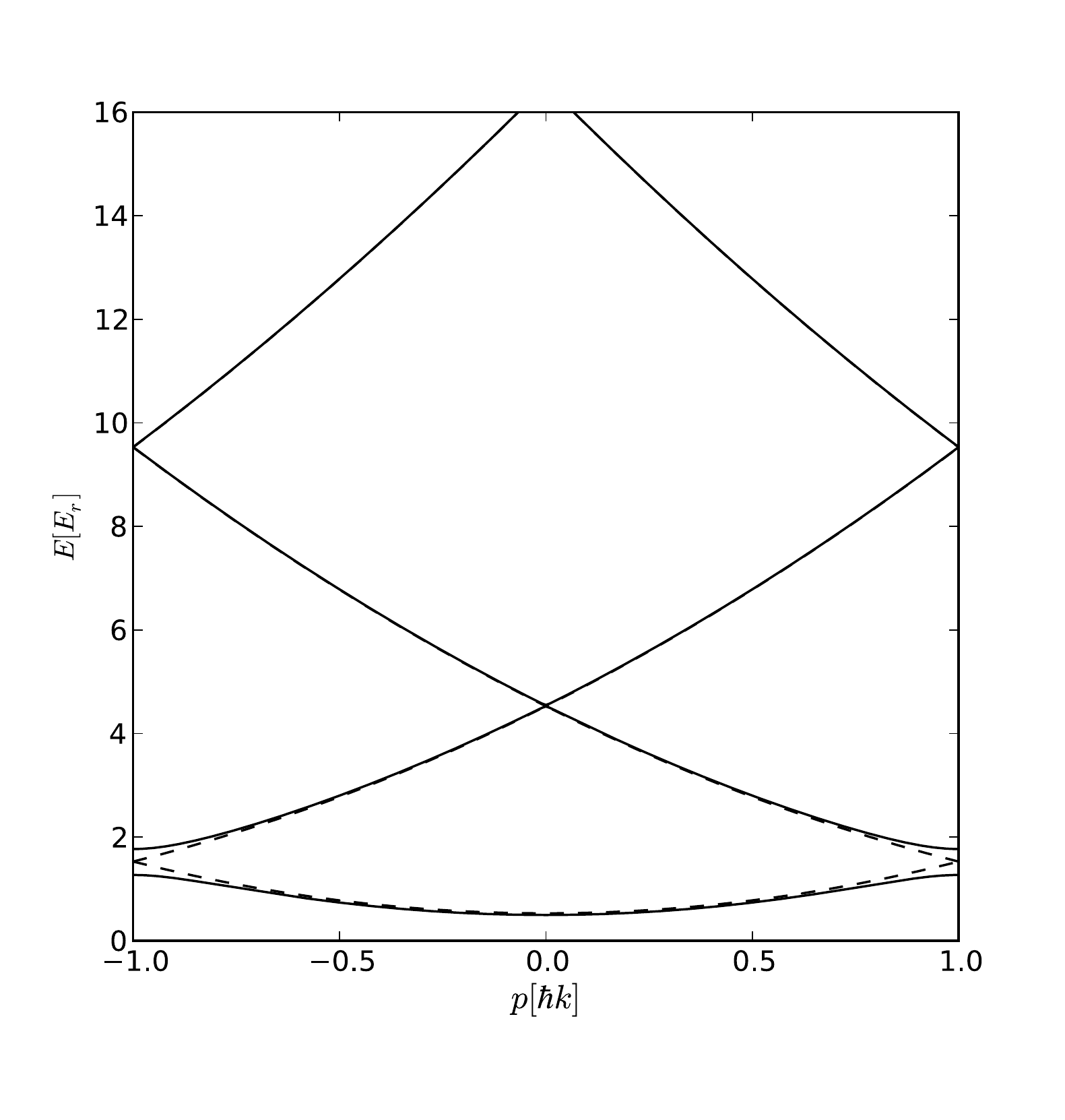}
\caption{\label{fig:band_structure} Band structure of the Bloch Hamiltonian. Left : tight binding limit ($U_0 = 10E_r$). Right : weak binding limit ($U_0 = E_r$).}
\end{figure}

There are two limits for which the eigenvalues can be analytically calculated : the weak binding limit and the tight binding limit. The tight binding limit is the regime where the particle is localized in the lattice site. In this case, the size of the wave function is smaller than the lattice step $\lambda/2$. Therefore, its momentum is larger than $2\hbar k$ and its kinetic energy larger than $4E_r$. This situation is possible only if the lattice depth is larger than the particle's kinetic energy, i.e. $U_0 \gg 4E_r$. In the tight binding limit, for low values of $n$, atoms are trapped in each well of the lattice and the energy $E_n(q)$ depends mainly on the state of the atom in the well (band index $n$) and almost not on the phase difference between two adjacent sites (quasi-momentum $q$) (see Fig.~\ref{fig:band_structure}). 

The other limit is the weak binding limit, where the optical lattice is a small perturbation to the free particle : $U_0 \ll 4E_r$. 
In this limit, the lattice couples the momentum eigenstates only at the edge of the first Brillouin zone ($q$ close to $1$). It couples the state $q-2$ and $q$, which have almost the same energy. The Hamiltonian is then the $2\times 2$ matrix : 
\begin{equation}
\begin{bmatrix}
q-2 & \kappa \\
\kappa & q
\end{bmatrix}
\end{equation}
that can be easily diagonalised. The energies are significantly shifted from the free energy when $|q-1| \leq \kappa$. There is a gap of width $2\kappa$ between the first and second band (see Fig.~\ref{fig:band_structure})

This condition $|q-1| \leq \kappa$ occurs when the Raman transition is resonant. In this case, one can transfer atoms from state \ket{-\hbar k} to state \ket{\hbar k}. This transition is called a Bragg transition : the atom is diffracted by the optical lattice. 

\section{Bloch oscillations}

\subsection{Adiabatic passage}

\begin{figure}
\begin{center}
\includegraphics[width=.49\linewidth]{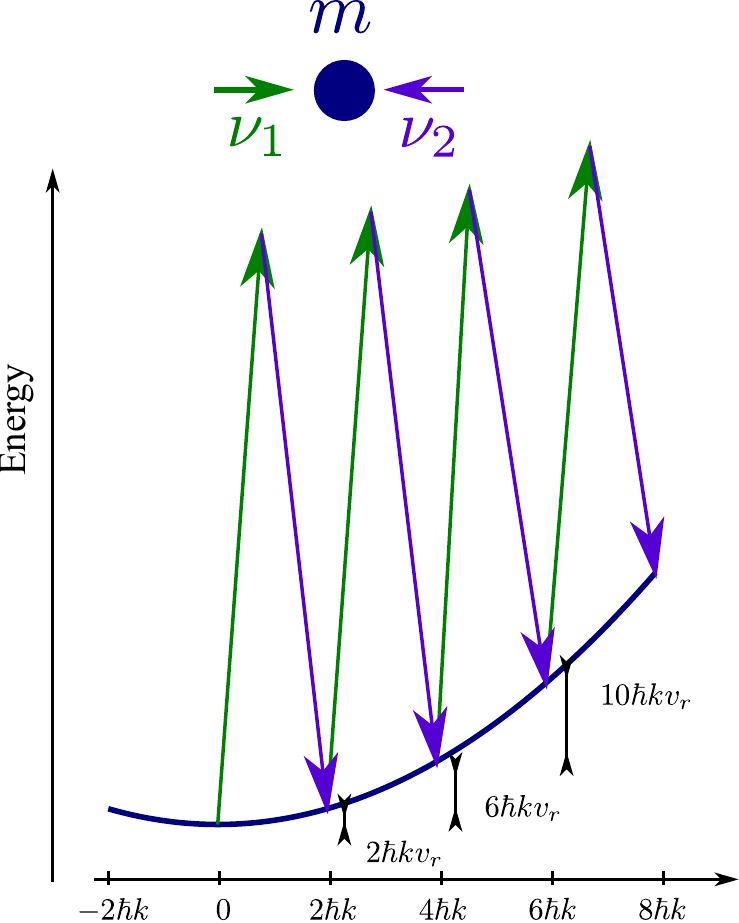}
\caption{\label{fig:BO} Bloch oscillation seen as a succession of Raman transition. On the x axis, the velocity of atoms and on the y-axis their energy. In
order to compensate for the Doppler effect, one need to change the frequency difference between the two lasers. In other words, the lattice is accelerated. }
\end{center}
\end{figure}

In the case of a Bragg transition, the population is transferred between two states by the mean of a resonant coupling (the lattice, see Fig.~\ref{fig:BO}). Another way can be used to transfer the population : the adiabatic passage. In this case, the coupling is switched on and then the system is swept across the resonance. To understand this phenomenon, let us use a simple two-level system with a coupling set to $\Omega$ and an energy difference set to $\delta$. 
The Hamiltonian is the following : 
\begin{equation}
H = \begin{bmatrix}
0 & \frac\Omega2 \\
\frac\Omega2 & \delta
\end{bmatrix}
\label{eq:hamil_adiab}
\end{equation}

\begin{figure}
\begin{center}
\includegraphics[width=.7\linewidth]{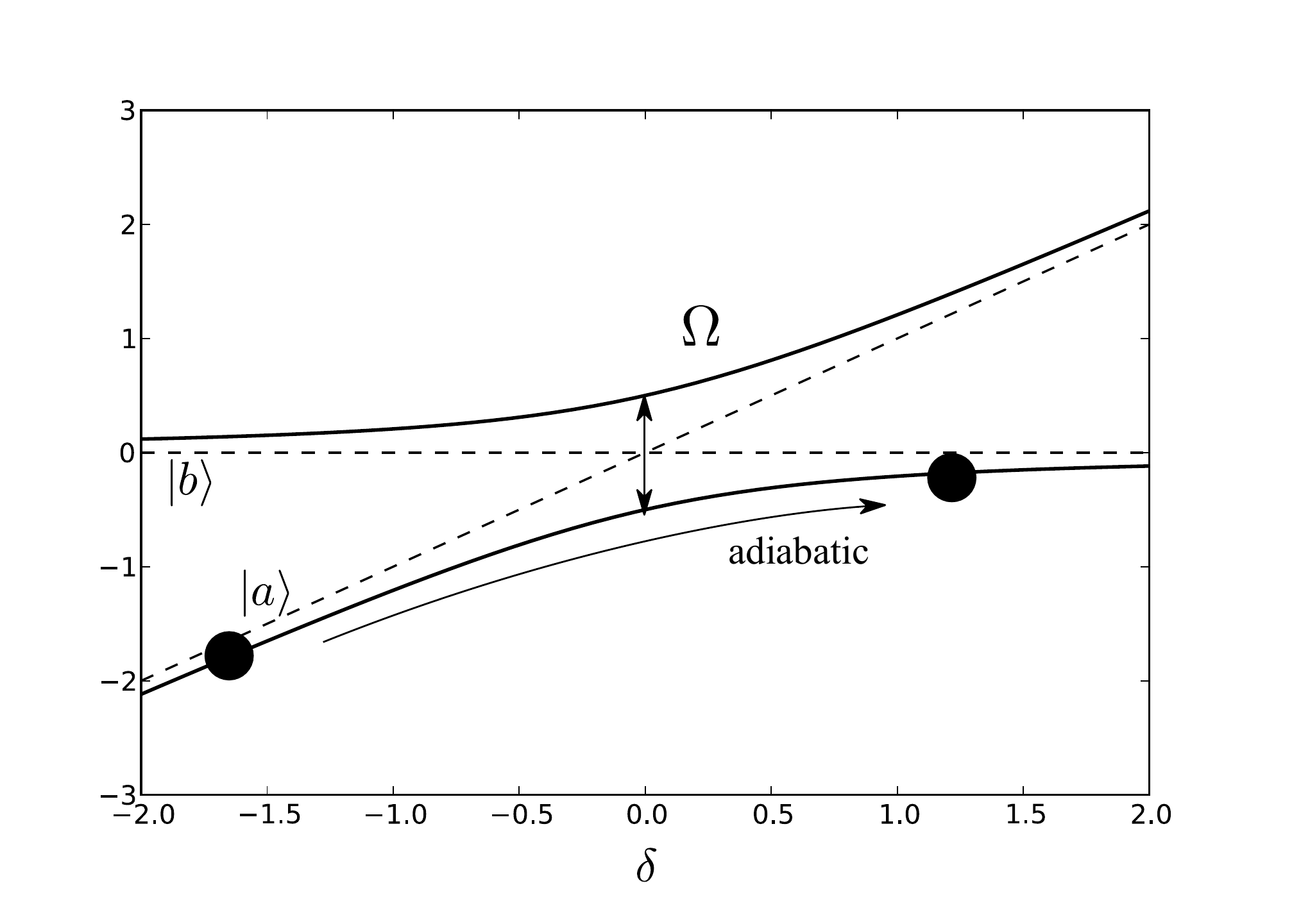}
\caption{\label{fig:adiabatic} Eigen-energies of the two-level system described by eq.~\ref{eq:hamil_adiab}. }
\end{center}
\end{figure}

We have depicted the eigenvalues of this Hamiltonian as a function of $\delta$ for a given $\Omega$ on fig.~\ref{fig:adiabatic}. By slowly changing the detuning $\delta$, the atom will adiabatically follow the eigenstates of the system and therefore will be transferred from state \ket{a} to state \ket{b}. This is true if the rate at which the eigenstates change is smaller that the coupling rate. The eigenstates are modified by the coupling when $\delta$ is on the order of $\Omega$. Therefore, for a constant change in $\delta$ ($\delta = \beta t$), the eigenstate will change over a typical time $\Omega / \beta$. The adiabatic criterion the says that we need $\beta \ll \Omega^2$.

\subsection{Bloch oscillations}
This two-level system can be used to understand first order Bragg transition. In order to sweep across the resonance of the transition, the velocity of the lattice is changed (the lattice is accelerated). Let us call $a$ this acceleration : $v(t) = a(t-t_0)$. The difference in energy between two states will evolve at a rate of $2ka$ and the criterion is therefore $a \ll \Omega^2/k$. Using the value of $\Omega$ as a function of the lattice depth $U_0$ and introducing the natural acceleration $a_0 = \hbar^2k^3/m^2$ , we obtain : 
\begin{equation}
a \ll \frac{a_0}{16}\left(\frac{U_0}{E_r}\right)^2
\end{equation}
For Rb atoms, the natural acceleration $a_0$ is about $260\unite{m/s^2}$. 

\begin{figure}
\includegraphics[width=.7\linewidth]{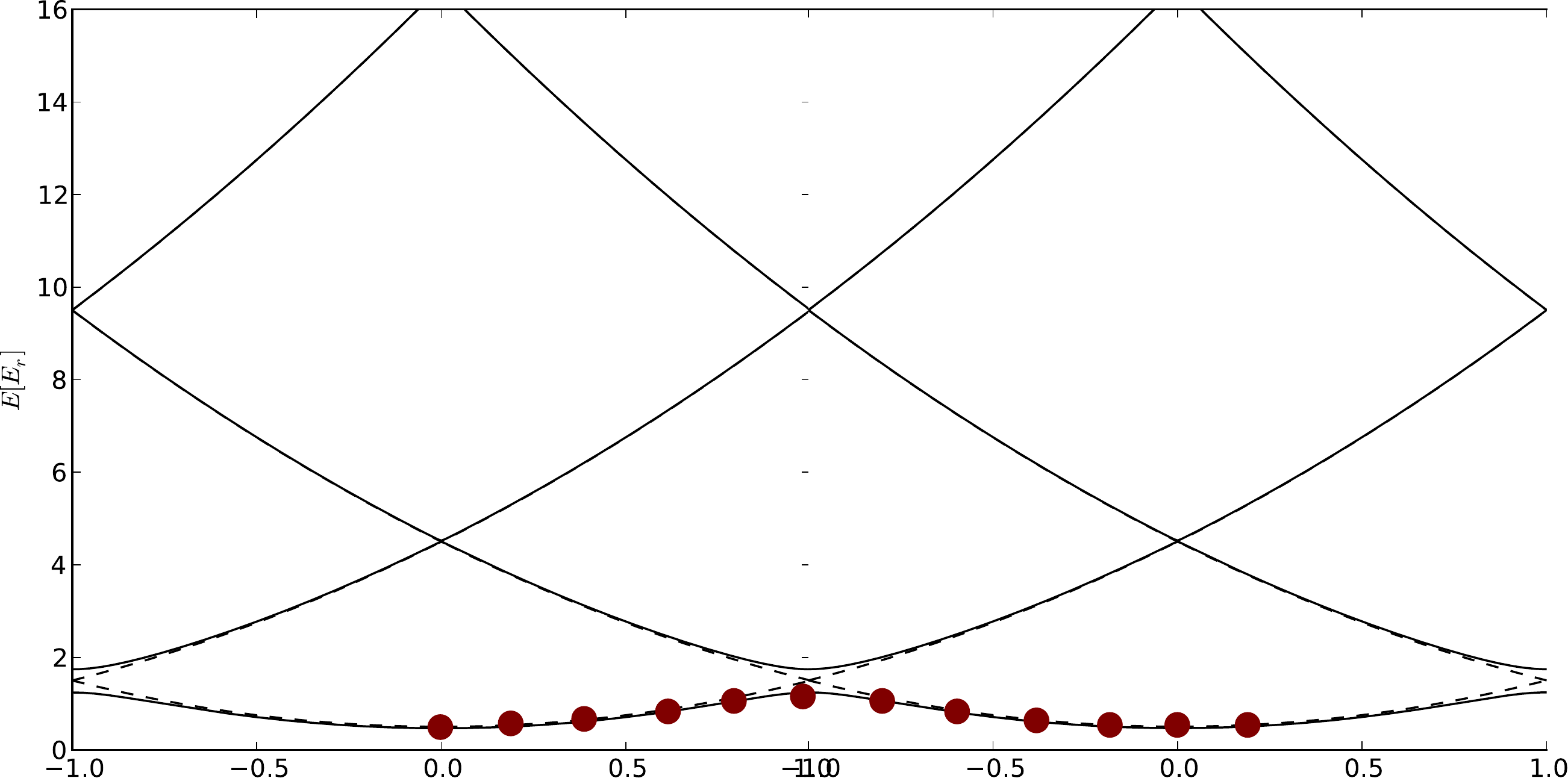}
\caption{\label{fig:Bloch_oscillation} Visualisation of a Bloch oscillation in the quasi-momentum representation. The x-axis represents the quasi-momentum. The best representation of the quasi-momentum is not $q$ directly but $e^{iqd}$, i.e. a point on a circle. This picture should be printed on a cylinder. Here, the x-axis represents two turns around the cylinder. In the adiabatic regime, after one oscillation, the state of the atom is unchanged}
\end{figure}

Fig.~\ref{fig:Bloch_oscillation} describes the full evolution of the eigenstate in the case of an adiabatic transition. The atomic state stays in the same band index $n$, and the quasi-momentum evolves continuously. The best representation of the quasi-momentum is a number on a circle; after one turn, the system comes back to the initial state. It oscillates. This is the Bloch oscillation. The period $T_\mathrm{Bloch}$ is set by the time in which the quasi-momentum increases by $2\hbar k$. In the case of a uniform acceleration $a$, we have 
\begin{equation}
T_\mathrm{Bloch} = \frac{2\hbar k}{ma}
\end{equation}

\subsection{Landau-Zener criterion}
There is a more quantitative approach to this problem due to Clarence Zener\footnote{Zener is well known for Zener diodes. As we will see in the next section, the problem of electrical breakdown in a diode was studied using the same Hamiltonian as for Bloch oscillations} \cite{Zener} and independently to Lev Landau. 

In the two-level system, the so-called Landau-Zener formula states that the non-adiabatic transfer probability is:
\begin{equation}
P = e^{-2\pi\frac{\Omega^2}{4\beta}}
\end{equation}

In the case of Bloch oscillations, we obtain a loss probability :
\begin{equation}
P_\mathrm{losses} = e^{-a_c/a}\ \ \mathrm{with}\ a_c = a_0\frac\pi{64}\left(\frac{U_O}{E_r}\right)^2
\end{equation}
For $U_0 = 4E_r$ and $a_c = 200\unite{m/s^2}$, losses due to gravity are $5\times 10^{-12}$ per oscillation in the case of Rb atoms. This proves how efficient Bloch oscillations are. This efficiency will not be degraded if there are fluctuations of the lattice depth (this is not the case for a Bragg transition, were the $\pi$ condition should be fulfilled). Furthermore, this efficiency is independent of the initial velocity if the initial state is in the first band - again this is a great difference between Bloch oscillations and Bragg diffraction (or Raman transition)\cite{Szigeti2012}.



\subsection{Bloch oscillations with a constant force}
\label{sec:BO_force}

Bloch oscillations were introduced by Zener in 1932\cite{Zener}. In this paper, they study the problem of an electron in a (periodic) crystal with a constant force (the electric field). This problem is similar to the problem that we solved in the previous paragraph: if we constantly accelerate the optical lattice, then in the frame of that lattice, there will be a constant inertial force. 

In Zener's problem, the Hamiltonian is the following : 
\begin{equation}
H = \frac {\hat{p}^2}{2m} -F\hat{x} + U(\hat{x})
\label{eq:sta}
\end{equation}
where $U(x) = \frac{U_0}2 \cos(2kx)$ is the periodic potential. This Hamiltonian is not invariant over a translation by $\lambda /2$, however it is still possible to calculate the evolution of the $T_d$ operator, in order to get the evolution of the quasi-momentum. 
\begin{eqnarray}
-i\hbar \dadt T_d &=& \left[H, T_d\right] \\
&=& \left[-F\hat x, T_d\right] \\
&=& FdT_d
\end{eqnarray}

Therefore, if we start with a state of quasimomentum (or momentum) $q(0)$, the state will remain an eigenstate of the translation operator, with an eigenvalue $q(t) = q(0) + Ft/\hbar$. We want the emphasize that this equation is true, what-ever the intensity of the force. Because the quasi-momentum is defined modulo $2k$, this linear behaviour is in fact a periodic one, with a period equal to $2\hbar k /F$.

The solution of the Schr\"odinger equation will be of the form $\psi(x,t) = e^{iq(t)x}u(x,t)$, with $u(x,t)$ being a periodic function. This function is a solution of the Schr\"odinger equation with the following Hamiltonian : 
\begin{equation}
\frac{(\hat p + \hbar q(t))^2}{2m} + U(\hat{x})
\end{equation}
This is the equation that was solved previously. 

\subsection{Understanding Bloch oscillations in the tight binding limit}
In the previous sections, we have described Bloch oscillations as a succession of 2-photon Raman/Bragg transitions. This picture is only valid in the weak binding limit (when $U_0 \ll 4E_R$). Indeed, in deeper lattices it is not possible to consider atoms as free particles that are coupled only at the edge of the Brillouin zone. For such lattices, eigenstates are always in a superposition of different plane waves. 

\begin{figure}
\centering
\includegraphics[width=.7\linewidth]{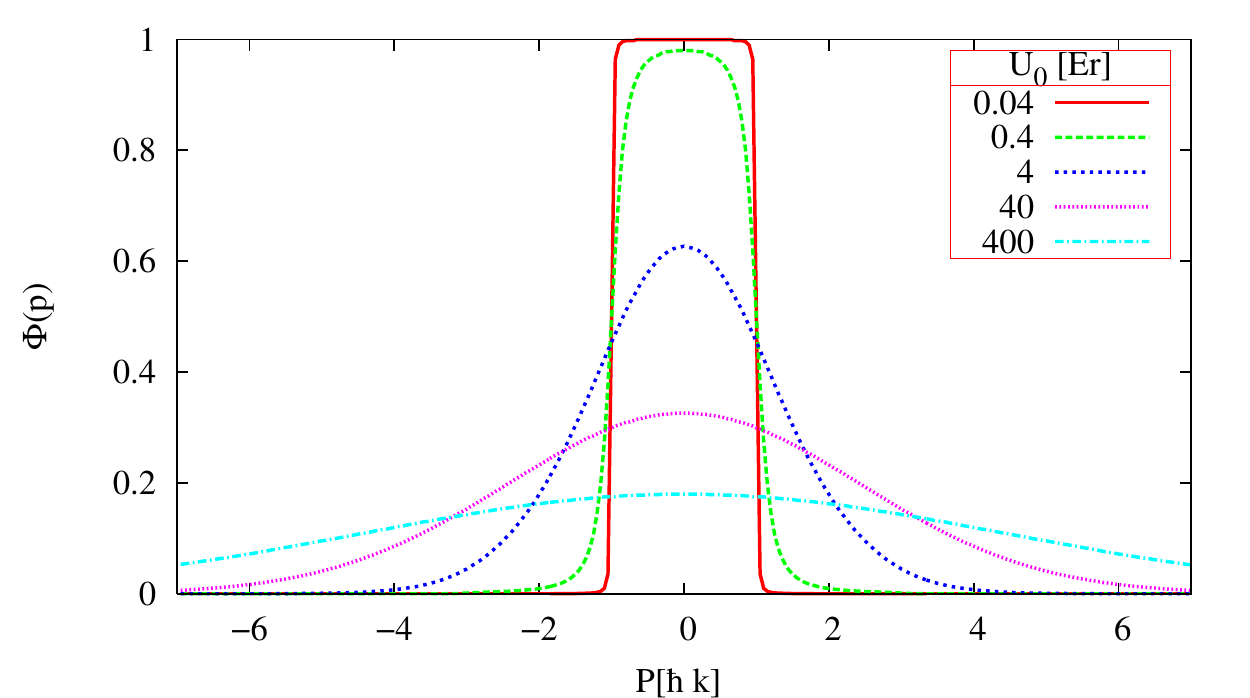}
\caption{\label{wannier} Wannier function in momentum space for different depth of the lattice}
\end{figure}

The Bloch theorem tells us that the eigenstate \ket{q,n} ($n$ is the band index and $q$ the quasimomentum) can be written as a superposition of plane waves $q + 2l\hbar k$ where $l\in \mathbb{Z}$. We can define the Wannier function \cite{Wannier}, in momentum space, as the coefficient $\Phi_n(p)$ such that : 
\begin{equation}
\ket{q,n} = \sum_l \Phi_n\left(q + 2l\hbar k\right) \ket{q + 2l\hbar k}
\end{equation}

This function depends on the depth of the lattice. We have plotted the Wannier function of the first band in fig.~\ref{wannier}. For very low lattice depths, the Wannier function is simply a square, because $\ket{q,n} \simeq \ket{q}$. In the tight binding limit, it is more convenient to think in the position space : atoms are trapped at the bottom of each lattice well and the Wannier function is simply the eigenstate of one lattice well. In the harmonic limit, it is a Gaussian with a width proportional to $1/\sqrt{\omega}$ where $\omega$ is the harmonic trapping frequency (which scales as $\sqrt{U_0}$). Therefore, in momentum space, the Wannier function is also a Gaussian with a width proportional to $U_0^{\frac14}$.

\begin{figure}
\centering
\includegraphics[width=.49\linewidth]{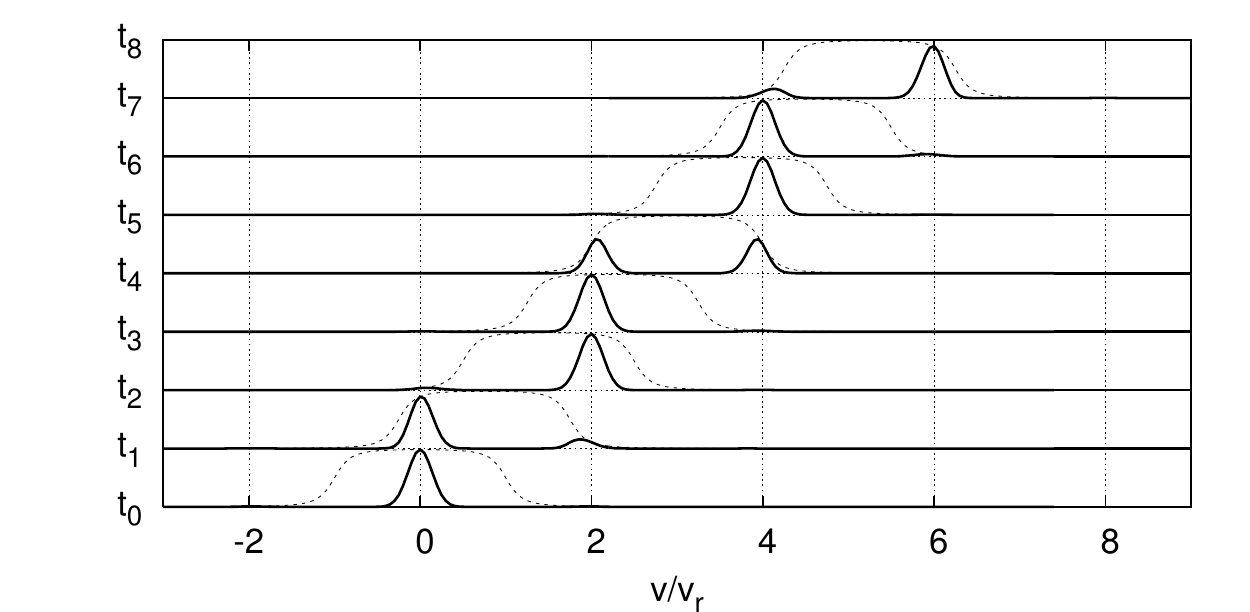}
\includegraphics[width=.49\linewidth]{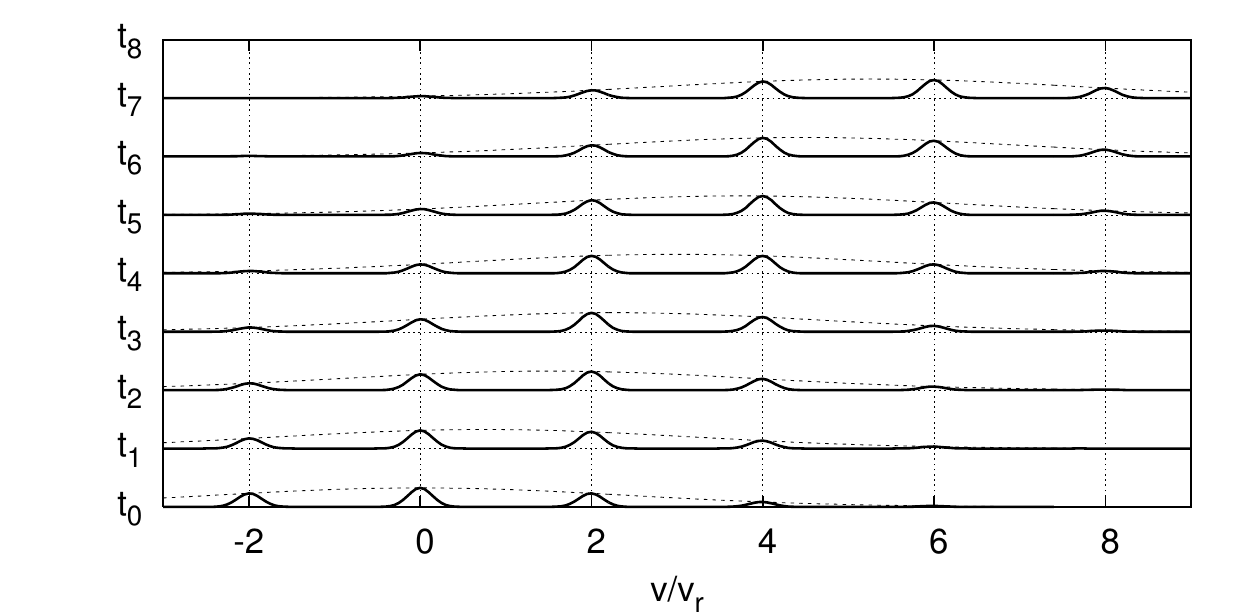}
\caption{\label{evolution} Description of the evolution of the velocity distribution (plain line) of atoms inside the lattice. This distribution if given by the Wannier function (dashed line). Left : weak binding limit ($U_0 = .4E_r$). Right: Tight binding limit ($U_0 = 40 E_r$).}
\end{figure}

\begin{figure}[t]
\begin{minipage}{.45\linewidth}
\includegraphics[width=.9\linewidth]{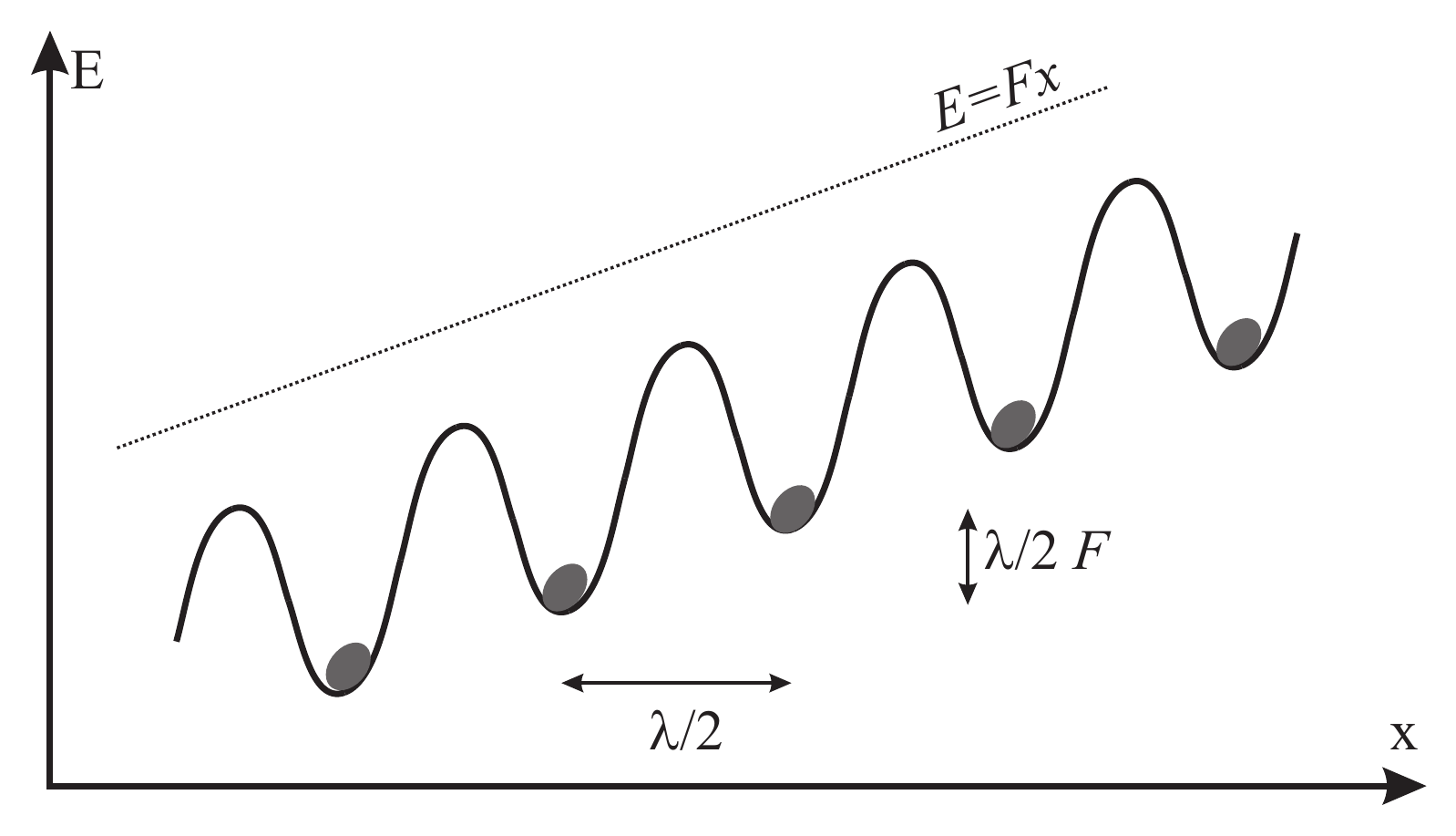}
\end{minipage}
\begin{minipage}{.45\linewidth}
\caption{\label{fig:energie_pot} Bloch oscillations in the tight binding limit. In this limit, atoms are trapped in the lattice. Because of the potential induced by the force, there is a phase shift between sites. This phase shift is a multiple of $2\pi$ for times that are multiple of the Bloch period .}
\end{minipage}
\end{figure}

The advantage of the Wannier formalism is that we have a continuous description of Bloch eigenstates in weak and tight binding limits. Using this formalism, it is easy to visualize Bloch oscillations. Figure~\ref{evolution} presents the momentum distribution of atoms in the fundamental band of an accelerated lattice. The Wannier function, which is constant in the frame of the lattice, is accelerated in the lab frame. The envelop function is shifted in momentum space while the quasi-momentum remains constant. In the weak binding limit, we observe that atoms have the same velocity and jump by steps of 2 recoil velocities (for example between instants $t_1$ and $t_2$ on the graph), except at the edge of the first Brillouin zone (instant $t_4$).

In the tight binding limit, the system evolves continuously and the mean velocity of atoms is the same as the velocity of the lattice. The true nature of the oscillation appears only when we let atoms between different sites interfere. This is done by adiabatically switching off the lattice. In the frame of the lattice, there is an intuitive way of understanding the Bloch oscillations. In this frame, atoms are submitted to a constant force. One can plot the potential energy given in equation \ref{eq:sta} (see Fig.~\ref{fig:energie_pot}). In the tight binding limit, atoms are trapped in the lattice. Their mean velocity is zero (the same as the velocity of the lattice). Because of the potential induced by the force, there is an energy shift between the bottom of each well. This energy is equal to $\frac\lambda2F$. This will induce a phase shift between two adjacent sites: $\delta\phi = \frac{\lambda Ft}{2\hbar}$. For instants multiples of the Bloch period, this phase shift is exactly equal to $2\pi$ and the system is in the initial state.

\afterpage{\clearpage}

\begin{center}
\Large\textbf Lecture II : application of Bloch oscillation in atomic physics
\end{center}

In the first section, we have seen advantages of Bloch oscillations. In particular, we have seen that they are not sensitive to the initial velocity of atoms (provided that it fits within the first Brillouin zone); they are not sensitive to variations of the intensity (provided that the intensity is above a given threshold) and they also work in the tight binding limit with a very good efficiency. 

The first observations of Bloch oscillation in atomic physics dates from the 1996. In Paris, the group of C.~Salomon managed to use Bloch oscillation to accelerate Cs atoms. They were able to precisely measure the evolution of the momentum distribution of atoms during the Bloch oscillations\cite{BenDahan}. They also demonstrate how efficient Bloch oscillations are for transferring a large number of recoils to atoms\cite{Peik}. The same year, the group M.~Raizen was also able to observe resonance at the Bloch frequency in an accelerated lattice\cite{Wilkinson}. Those preliminary experiment have opened new perspectives for high precision measurement and atom interferometry. In this lecture, I will focus on the three main application of Bloch oscillations in this field : 
\begin{itemize}
\item Gravity measurement
\item Large Momentum Transfer Beamsplitters using Bloch oscillations
\item Measurement of the recoil velocity
\end{itemize}

\section{Measurement of the acceleration of gravity}

In this section, we will briefly describe how Bloch oscillations can be used in an atomic gravimeter. This section is a short review of the different ways of using Bloch oscillations to measure gravity and a more precise description of a recent experiment that we have made at the Laboratoire Kastler Brossel in Paris. The lecture of G.~Tino of this school gives details of his experiment \cite{Tino2013} and the lecture of P.~Bouyer gives the state of the art in the field of gravimetry using cold atoms \cite{Bouyer2013}. 

In section \ref{sec:BO_force} of this lecture, we have described the behaviour of atoms in an optical lattice that are submitted to a constant force $F$. When the depth of the lattice is high enough, the atomic wave-function will have a periodic behaviour with a period $T_\mathrm{Bloch}$ given by : 
\begin{equation}
T_\mathrm{Bloch} = \frac{2\hbar k}{F}
\end{equation}
A measurement of this period can therefore be used to measure the force on atoms. 

There is a periodic behaviour of both the position and the velocity of the atomic wave-packet. Any measurement, using either the position, the velocity or a combination of both (using the time-of-flight technique) will exhibit this periodicity.

The first measurement of $g$ using Bloch oscillations comes from the group of M.~Inguscio in 2004 \cite{Roati}. They use a fermion in order to reduce decoherence. Using about 200 BOs, they reach a sensitivity on the order of $10^{-4}$. In 2006, the group of G.~Tino, using bosonic Sr, which has very small atom-atom interactions, managed to observe more than 5\,000 oscillations\cite{Ferrari2006}. The sensitivity in $g$ is $5\times 10^{-6}$. In both experiments, a very small and dense cloud of atoms was used. Despite the poor resolution, the main advantage is that the gravity is measured with a very small probe (on the order of $100~\mu m$) and with a probe which is almost not moving (the motion inside the lattice is in the $\mu m$ range). This opens new methods for measuring forces at small distances \cite{Carusotto}. In such an experiment, because the position or velocity measured with the time-of-flight technique is not very precise, the accuracy comes from the possibility to measure a high number of oscillations. However, because of atom-atom interactions, there is a decoherence that limit the number of Bloch oscillation and special care should be taken. We should also note the work done in the group of H.~C.~Nagerl on Cs where a Feshbach resonance was used to cancel atom-atom interactions and more than 20\,000 oscillations were observed \cite{Gustavsson2008}.

In the previous paragraph, we have presented experiments where the oscillations were directly observed. Another set of experiments should be mentioned, where this periodic phenomenon is indirectly observed. The idea is to excite the system at the Bloch frequency and see a resonance peak. In the group of G.~Tino, working on Sr atoms, the amplitude of the optical lattice is modulated at a frequency close to the Bloch frequency (or a multiple of this frequency). They are able to see the resonance by observing the width of the cloud:  this width increases at resonance because of resonant tunnelling between adjacent sites. Using this method they are able to measure gravity with an uncertainty of $1.5\times 10^{-7}$ over one hour \cite{Poli2011}. A more precise description of this experiment, as well as recent results, is given in G.~Tino's lecture notes of this school. Another precise experiment, aiming at measuring forces at small distance has been developed by F.~Pereira Dos~Santos at SYRTE in Paris. The idea is to optically measure the energy shift between sites in the lattice using Raman transitions\cite{Wolf2007}. They have recently demonstrated a relative sensitivity of $10^{-5}$ at 1~s \cite{PhysRevA.87.023601}. 

Bloch oscillations can also be used in gravity measurement in a different way. As we have seen, each Bloch oscillation transfers to atoms a well-known momentum $2\hbar k$. Using Bloch oscillations in an atomic gravimeter that measures the acceleration of atoms, it is possible to compensate the free fall of atoms. A longer integration time can then be used without the disadvantage of a large trajectory. The first precise measurement was carried in the group of F.~Biraben at LKB \cite{Clade2} with 50 oscillations of rubidium atoms. 

\begin{figure}
\begin{center}
\includegraphics[width=.5\linewidth]{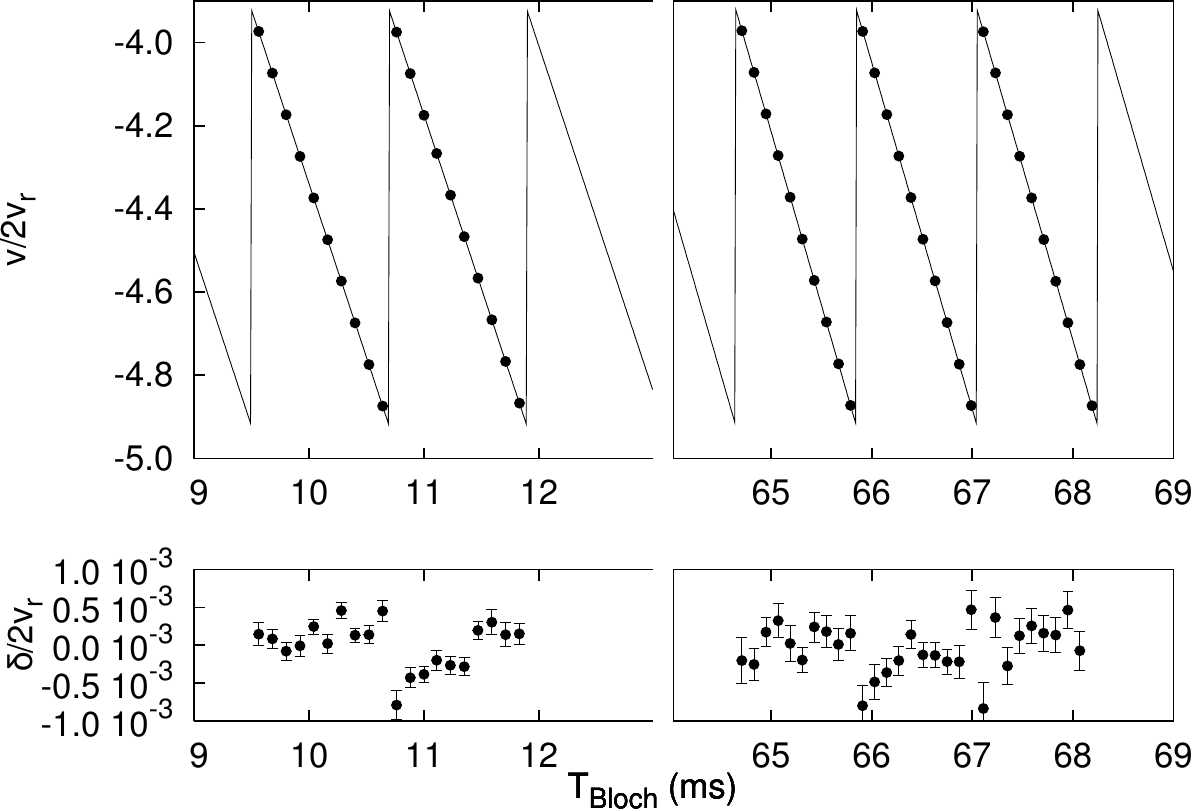}
\end{center}
\caption{\label{fig:BlochGravite} The center of the final velocity distribution vs. the duration of the standing wave. The dots represent the experimental data and the line the least-square fit performed by fixing the recoil velocity.}
\end{figure}

In this experiment, the velocity of atoms after the Bloch oscillations is given both by the gravity $g$ and the number $N$ of oscillations : 
\begin{equation}
v(t) = gt - 2N\frac{\hbar k}m = g(t - NT_\mathrm{Bloch})
\end{equation}
where $T_\mathrm{Bloch} = \frac{2\hbar k}{mg}$ is the Bloch period.
Because the integer number $N$ increases by steps and is roughly equal to $t/T_\mathrm{Bloch}$, the velocity evolves in a sawtooth shape (see Fig.~\ref{fig:BlochGravite}). 

A velocity sensor using Doppler-sensitive Raman transitions was used to precisely measure the velocity of atoms. The number of oscillations was then limited by the residual Rb vapour in the chamber.  We were able to reach a precision of $10^{-6}$ over one hour of integration. 

This was the first time that Bloch oscillations were used in combination with a precise velocity sensor. The idea is to be able to take advantage of the most precise velocity sensor -- but keeping the overall motion of atoms in a small region compared to other atom gravimeter. In the group of A.~Bresson at ONERA, they adapt this method to atom interferometry \cite{PhysRevA.85.013639} using a sequence similar to the sequence used for our measurement of $h/m$ (this sequence will be described in section \ref{sec:mes_scheme}, Fig.~\ref{fig:sequencehsurmB}): they use gravity-induced Bloch oscillations between the two sequences of $\pi/2$ pulses. They reach a sensitivity of $2\times 10^{-7}$ in 300~s. To obtain such a precision, they are using 70 BOs in a stationary vertical standing wave.

\begin{figure}
\includegraphics[width=.49\linewidth]{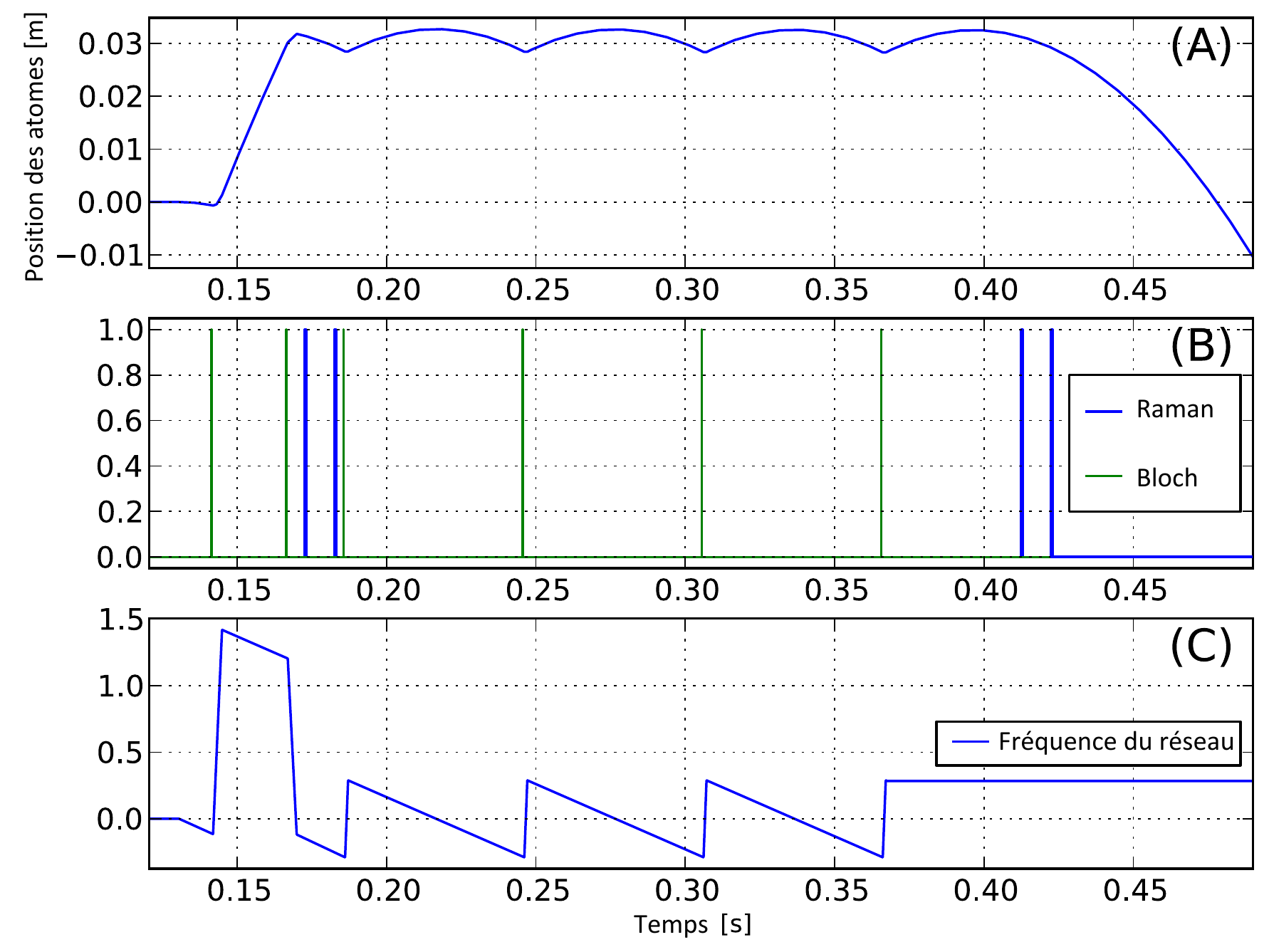}
\includegraphics[width=.49\linewidth]{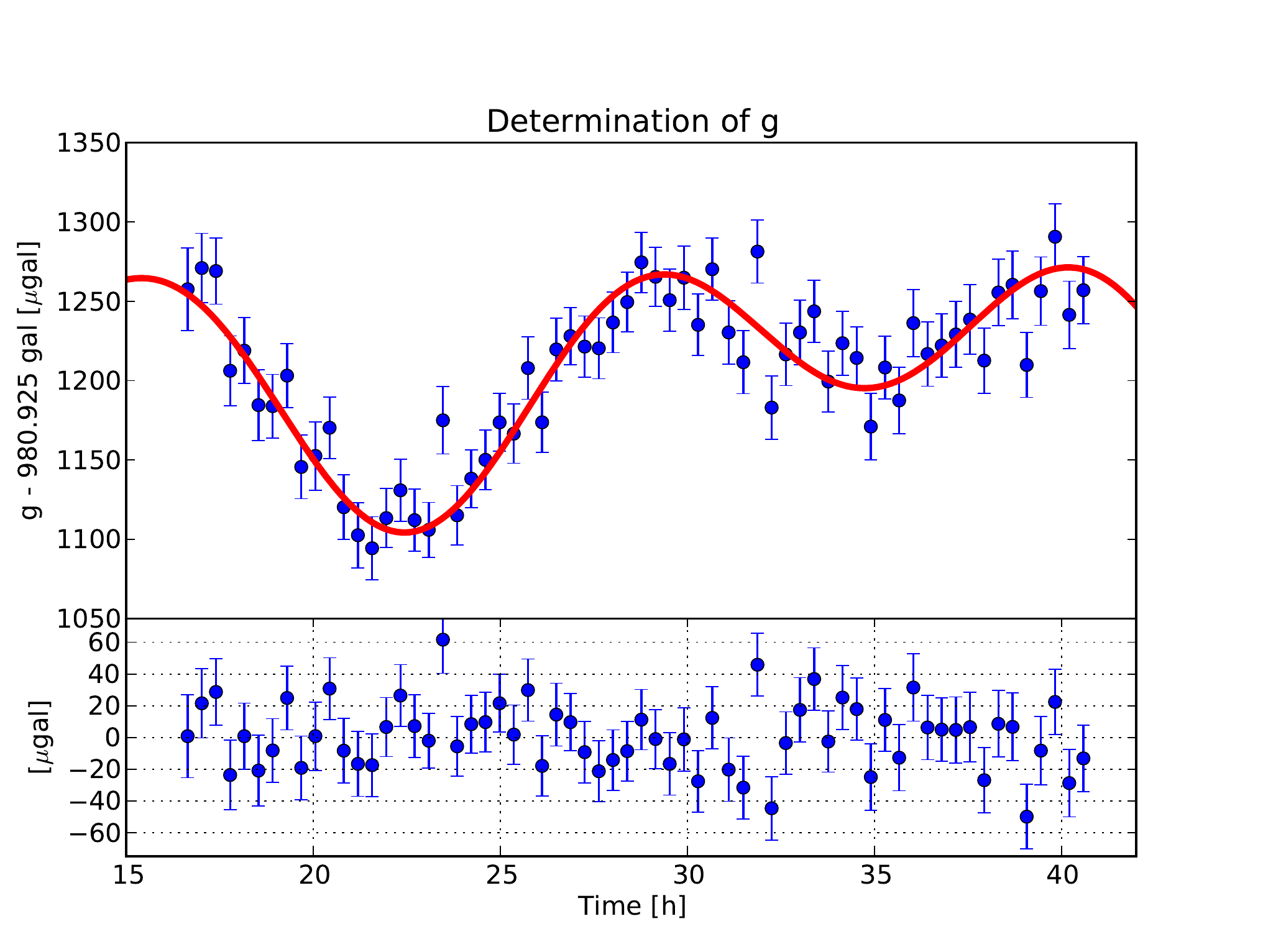}
\caption{\label{fig:gravi_bloch_pulse} Left : (A): trajectories of the atoms during the measurement
procedure. The atoms are held at a given position using an
atomic elevator, we apply the first pair of $\pi/2$ Raman pulses,
then they are maintained against gravity during 230 ms by
periodically transferring them 100 photon momenta.
(B): timing sequence of the Raman and Bloch beams, the two
first Bloch pulses are used to perform the atomic elevator.
(C): the frequency of the optical lattice versus time. Right :Gravity data taken over one day fitted by the Earth-tide model. Each data point is deduced from the average over
6 measurements. The lower curve shows residuals of the fit
($1 \unite{\mu Gal}=10^{-8} \unite{m/s^2}$). } 
\end{figure}

Recently we have presented a new scheme of compact atomic gravimeter based on atom interferometry\cite{PhysRevA.88.031605}. Atoms are maintained against gravity using a sequence of coherent accelerations performed by the Bloch oscillations technique. In order to compensate the fall of atoms between the pulses, we use a sequence of brief and strong accelerations based on Bloch oscillations in an accelerated optical lattice. Because the lattice is pulsed we have less decoherence compared to previously described methods where the force is applied continuously. In this sequence, the overall motion of atoms is about 5~mm and the precision we get is $4.8\times 10^{-8}$ with 4 minutes of integration time. Even if the precision is far from the state of the art in atomic gravimetry, this is the most precise measurement of gravity on a falling scale which is less than 1~cm. We are able for example to see the tidal effect on gravity (see Fig.~\ref{fig:gravi_bloch_pulse}). Furthermore, in the experiment, the position of atoms is precisely controlled using an "atomic elevator" based on Bloch oscillations \cite{Cadoret2008}. This method can therefore be used to make a precise gradiometer. 

\section{Large momentum transfer beamsplitter using Bloch oscillation}

\subsection{Beamsplitters in atom interferometers}
In order to improve the sensitivity of an interferometer, one can increase the duration $T$ of the interferometer. However, this is not always possible, as it will 
also require a larger experimental setup (the size of the setup will scale as $T^2$ for a free fall gravimeter). 

Instead of increasing $T$, it is possible to use a beamsplitter with a higher momentum transfer (Large Momentum Transfer beamsplitter, LMT). If the beamsplitter transfers $2M$ recoil momenta to the atoms instead of $2$, then the accuracy is improved by a factor $M$. This possibility to improve the sensitivity of interferometers raised interest among physicists. 

In the measurement of the recoil velocity experiment made in Paris, the duration of the interferometer is limited by the size of the vacuum chamber. Longer durations could be used with a larger cell but would require to control some systematic effects (mainly wave fronts and magnetic field) on a larger scale. The solution of LMT beamsplitters will improve the sensitivity without this constraint. In this experiment, one of the main systematics comes from the finite waist of the laser beams. Because beams are not perfect plane waves, the recoil transferred is smaller than $\hbar k$. By using a LMT beamsplitter, one can keep the same sensitivity, but using beams with a larger waist where this systematic effect will be reduced.

While LMT beamsplitters will be a great tool in atom interferometers, up to now, there are only proof-of-concept experiments and no precision measurement based on such a method were published.
Many different approaches have been proposed and have been demonstrated. LMT beam splitters have been achieved by applying sequential two-photon Raman transitions \cite{McGuirk2000} and by alternatively applying a single multiphoton Bragg diffraction \cite{muller:180405}.

In the group of Kasevich at Stanford, $102$ recoils were achieved on a Rb BEC \cite{Chiow2011}. The LMT is based on a sequence of Bragg diffraction. This experiment is still in a preliminary stage as no fringes are displayed but only correlation in the noise of two simultaneous interferometers. Recent results are discussed in the lecture of M.~Kasevich \cite{Kasevich2013}.

\subsection{LMT beamsplitter based on Bloch oscillations}

An LMT beamsplitter, based on Bloch oscillations, has been suggested in \cite{Denschlag}. The technique of Bloch oscillations\cite{Peik} has proven to be very effective to coherently accelerate atoms. 
It should therefore  allow to increase the momentum transfer in LMT beamsplitters. In 2009, we have demonstrated such a beamsplitter using laser cooled atoms\cite{clade_PRL2009}. A similar experiment that has been done at Stanford/Berkeley has reached $24$ recoils on their interferometer \cite{muller2009_PRL240403}. A strong reduction of the fringes contrast has affected both experiments. We have conducted a theoretical modelization of the experiment \cite{Clade2010_LMT} to understand the losses of contrast. The Bloch-based LMT has been recently implemented in the group of J.~Close at ANU\cite{PhysRevA.88.053620}. In an optically guided atom interferometer, they manage to obtain a $80\hbar k$ momentum separation. This is the state of the art of momentum separation using Bloch oscillations.

\begin{figure}
\begin{minipage}{.5\linewidth}
\begin{center}
\includegraphics[width = .8\linewidth]{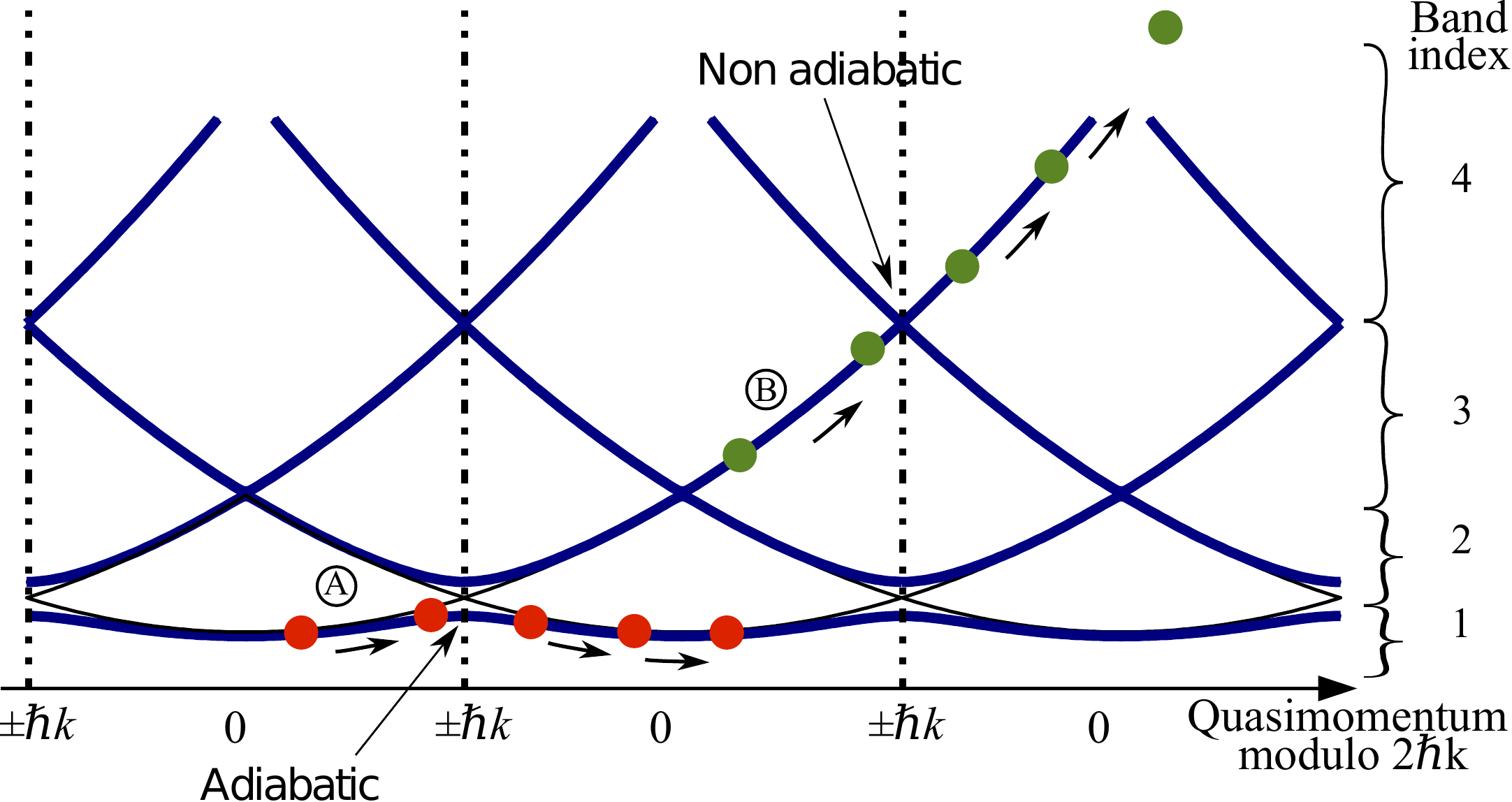}
\end{center}
\end{minipage}
\begin{minipage}{.4\linewidth}
\caption{Band structure of the optical lattice. Trajectories of the accelerated (A) and non accelerated atoms (B).}
\label{fig:bande}
\end{minipage}
\end{figure}

The principle of the large momentum transfer beam splitter consists in creating a superposition of two wave-packets separated by two recoil velocities and in loading them into the optical lattice such that one wave-packet is in the first band (A, see Fig.~\ref{fig:bande}) and the second one in the third band (B). A constant acceleration is then applied to the lattice. This acceleration (which acts like a force in the frame of the lattice) brings atoms to the edge of the first Brillouin zone. The acceleration of the lattice is small enough so that atoms in the first band have a large probability to make an adiabatic transition but high enough so that atoms in the third band change band. Each oscillation increases the momentum of the atoms by $2\hbar k$. On the other hand, atoms in the third band (atoms which change band) are not accelerated.

\begin{figure}
\begin{minipage}{.49\linewidth}
\includegraphics[width=\linewidth]{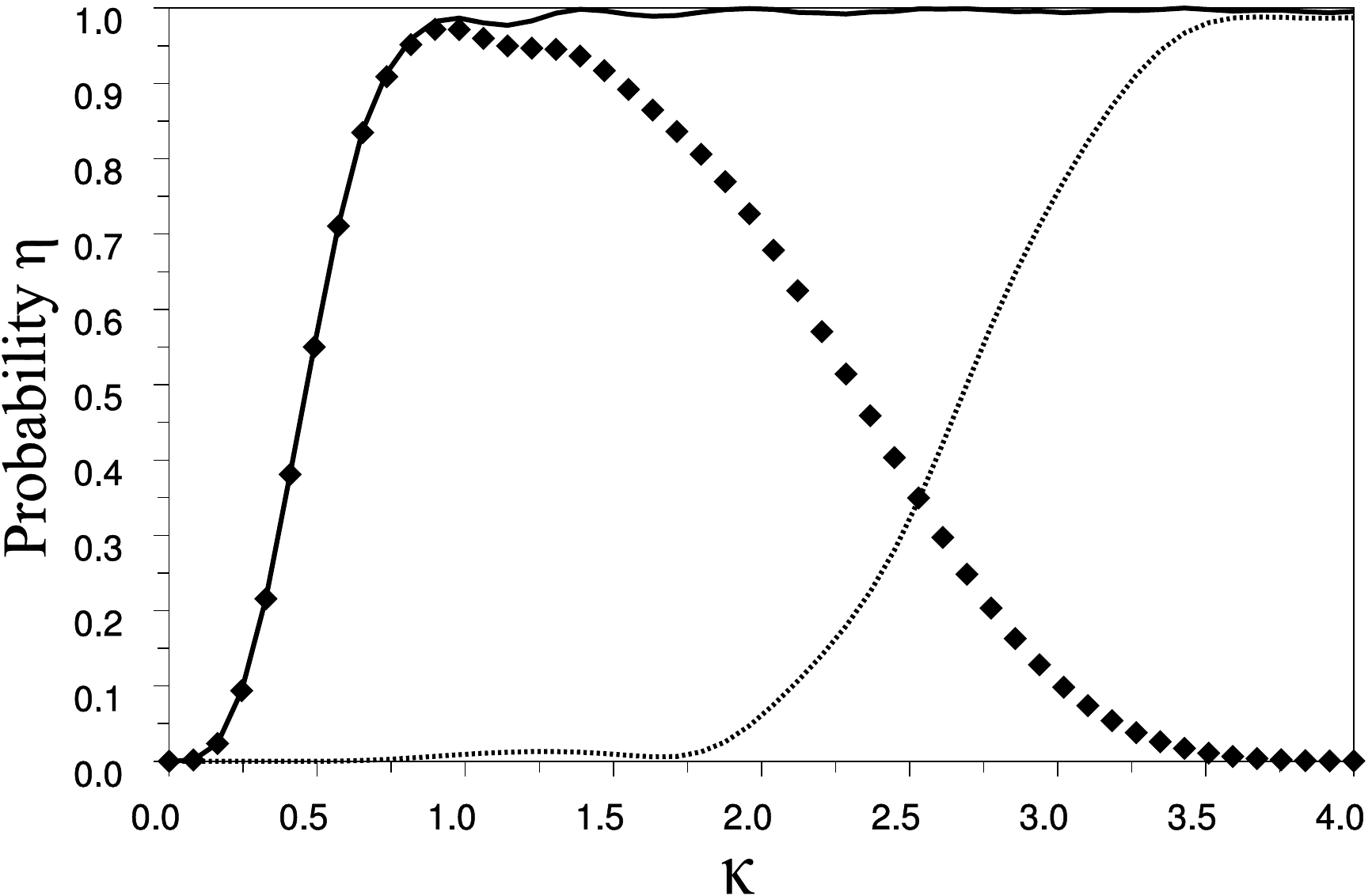}
\end{minipage}
\begin{minipage}{.49\linewidth}
\caption{\label{fig:LMT_proba}Transfer probability as a function of the maximal optical depth $\kappa = U_0/8E_r$ of the lattice. The acceleration is in 200\unite{\mu s} for N=2 recoils. Solid line: transfer probability for the first band $\eta_{11}$; dashed line: for the third band, $\eta_{33} \approx 1-\eta_{34}$; diamond: Efficiency of the LMT pulse, $\eta= \eta_{11}\eta_{34}$.} 
\end{minipage}
\end{figure}

The condition for the acceleration can be precisely calculated\cite{Clade2010_LMT}. We use the model described at the beginning of this lecture. We numerically integrate the full Hamiltonian for the two different initial conditions (atoms in the first or third Brillouin zone). Depending on the acceleration, we are able to calculate the probability to perform one Bloch oscillation (adiabatic criterion). This probability is plotted on Fig.~\ref{fig:LMT_proba}. The efficiency of the LMT beamsplitter is then the product of the probability to have one Bloch oscillation for the first band by the probability to not have BO for the third band. 

\begin{figure}
\begin{minipage}{.49\linewidth}
\includegraphics[width=\linewidth]{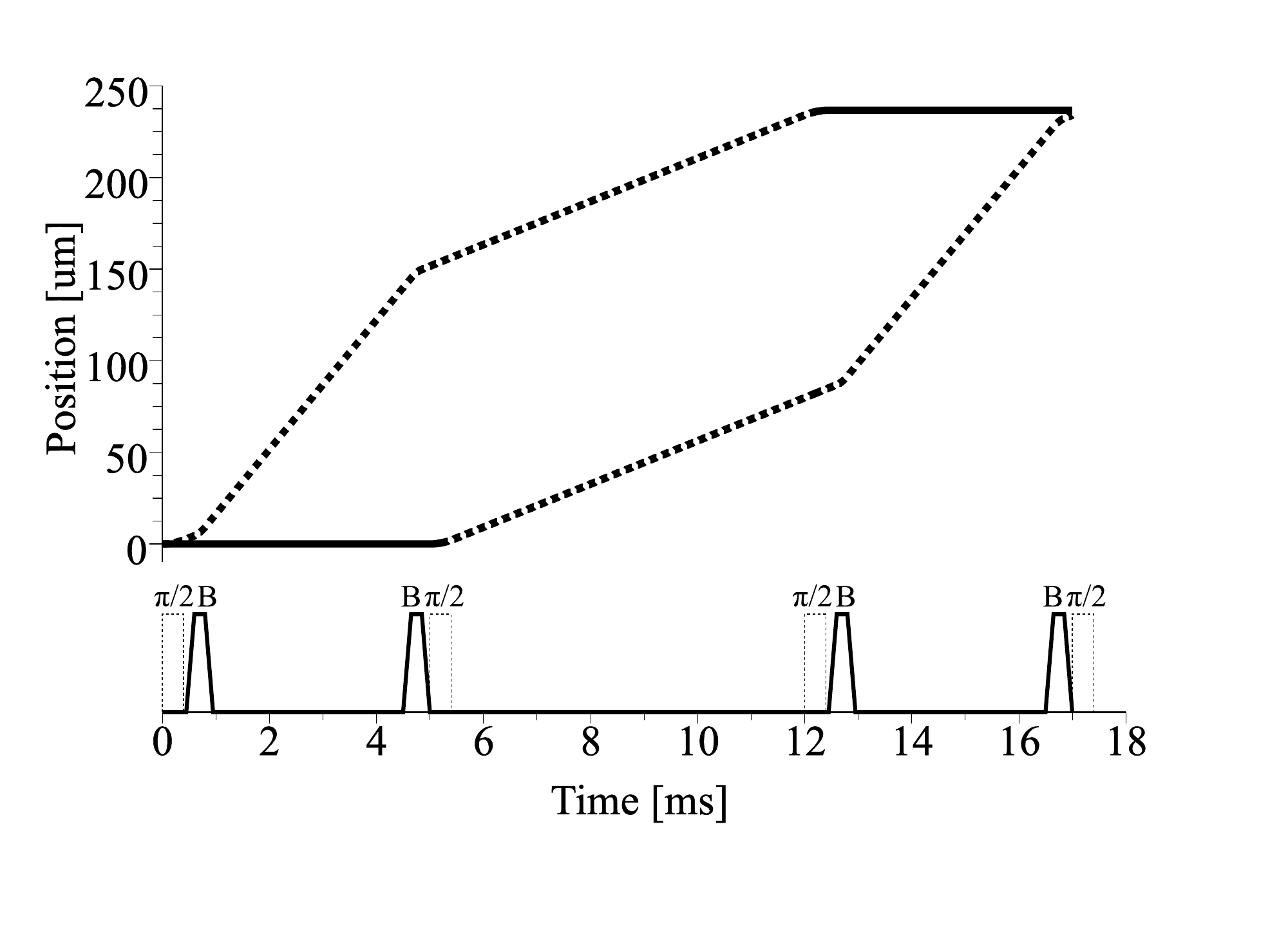}
\end{minipage}
\begin{minipage}{.49\linewidth}
\caption{\label{fig:LMT_schema} Schematic of the LMT based atom interferometer. This scheme is based on atom interferometer based on four $\pi/2$ pulses. Each pulse is followed (or preceded) by a Bloch acceleration. This configuration is symmetric in the sense that each atom spends the same amount of time in each internal state and Bloch state. Therefore at first order, shifts induced by the lattice are compensated and only the variations of such shifts appear.}
\end{minipage}
\end{figure}

A simple scheme for the interferometer is presented on Fig.~\ref{fig:LMT_schema}.  This scheme is based on an atom interferometer based with four $\pi/2$ pulses. Each pulse is followed (or preceded) by a Bloch acceleration. This configuration is symmetric in the sense that each atom spends the same amount of time in each internal state and Bloch state. Therefore at first order, shifts induced by the lattice are compensated and only the variations of such shifts are relevant. 

In order to understand the noise and the reduction of contrast in the experiment performed in 2009, we have conducted theoretical simulations of the interferometer\cite{Clade2010_LMT}. Using the same model, we are able to calculate precisely the phase shift of the interferometer based on Bloch oscillation. This phase shift has two origins: the kinematic phase due to the position of the lattice and the energy shift of atoms inside the lattice. 

The energy of the atoms is calculated as a function of their Bloch state and the depth of the lattice (we calculate the phase of the interferometer in the momentum space \cite{Schleich2013}). The phase shift between the two arms is then calculated by integrating the energy over time for the two paths of the interferometer. 

We impute this reduction to the phase shift induced by the lattice to the atoms. This problem in the BO-LMT can be understood by looking at Fig.~\ref{fig:bande}. The band gap between the first and second bands depends on the depth of the lattice. Therefore, an atom oscillating in the first band will endure a phase shift that depends upon the lattice depth. On the other hand, an atom in the excited band is insensitive to the lattice depth. There is therefore a differential phase shift that is accumulated between the two arms of the interferometer. 

This effect can be compensated using a symmetric configuration, but only if the intensity seen by the atoms remains constant along the interferometer which implies that the transverse motion of atoms is small. The conclusion of this theoretical study is that we need to use a colder atomic source. We are currently building a new experimental setup. Instead of using only laser cooling to prepare the atoms, a dipole trap will be used to cool atoms by evaporation. The aim of this project is first to increase the statistics on atom interferometer, and then, to reduce systematic effects. 

\section{Application of Bloch oscillations to the determination of the recoil velocity}

In this section, I will describe the experiment which is currently used to measure the atom recoil velocity at the level of $1.3\times 10^{-9}$ at LKB. 

\subsection{Raman transition, Doppler effect and atom interferometry}

In the first section, we have seen that two-photon Raman transitions can be used to transfer atoms between two internal states. In the case of counter-propagating transitions, there is a transfer of two photon recoils to the atoms. As we have seen this appears together with the Doppler effect. 

Applying a $\pi$ pulse to a cloud of atoms using a counter-propagating transition will result in the transfer of only the resonant atoms, i.e. those with the right velocity. Such a transition can be used to select a given velocity distribution. Its position will be determined by the frequency of the transition, and its width, by the strength of the transition. For a $\pi$ pulse, the velocity width will be given by $\Delta v \propto \frac{\Omega}{2k}\simeq \frac\lambda{4\tau}$ where $\Omega$ is the effective Rabi frequency and $\tau$ the duration of a $\pi$ pulse. 

\begin{figure}[b]
\begin{center}
\includegraphics[height=.3\linewidth]{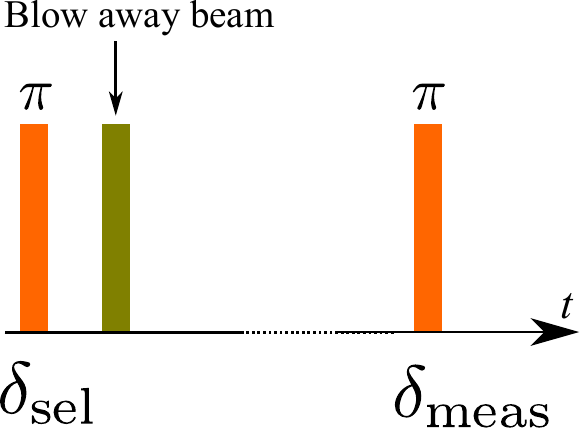}
\end{center}
\caption{\label{fig:seq_raman_simple} Description of velocity sensor based on a sequence of two $\pi$ pulses}
\end{figure}

A sequence of two $\pi$ pulses can be used to build an inertial sensor : the first $\pi$ pulse will select a velocity distribution (non-selected atoms are then eliminated). Then a second $\pi$ pulse is used to transfer atoms back to the initial state. If the velocity of atoms has changed between the pulses, then atoms are out of resonance and in order to be resonant, one needs to change the frequency of the second Raman transition ($\delta_\mathrm{meas}$) with respect to the frequency $\delta_\mathrm{sel}$ of the selection. More precisely, the final velocity distribution is measured by making different measurements at different frequencies $\delta_\mathrm{meas}$ and recording the proportion of atoms transferred (see Fig.~\ref{fig:seq_raman_simple}).

We can define the sensitivity in velocity of such a measurement as the inverse ($1/\Delta v$) of the velocity change that induces a significant decrease of the signal. In this experiment, the sensitivity is proportional to $\frac{4\tau}\lambda$. For example, in the case of an accelerometer, the sensitivity in acceleration will then be given by $\frac{4\tau T}\lambda$ where $T$ is the delay between the $\pi$ pulses. 

We should note that the precision at which a measurement can be performed depends both on the sensitivity and the signal-to-noise ratio. In the scheme described above, the sensitivity is better for a longer pulse duration $\tau$ (i.e. weaker Raman coupling). However, the signal (number of atoms selected) will also decrease and the signal-to-noise ratio will decrease. There is therefore no direct gain in reducing the Raman coupling if the dominant noise comes from the amplitude of the signal.

\subsection{Atom interferometry}

In the previous section, we have described an inertial sensor based on a sequence of two Raman $\pi$ pulses at two frequencies ( $\delta_\mathrm{sel}$ and  $\delta_\mathrm{meas}$ ). In this configuration, the velocity is measured as a Doppler effect seen in the transition probability of the second Raman pulse. 

\begin{figure}
\begin{center}
\includegraphics[height=.3\linewidth]{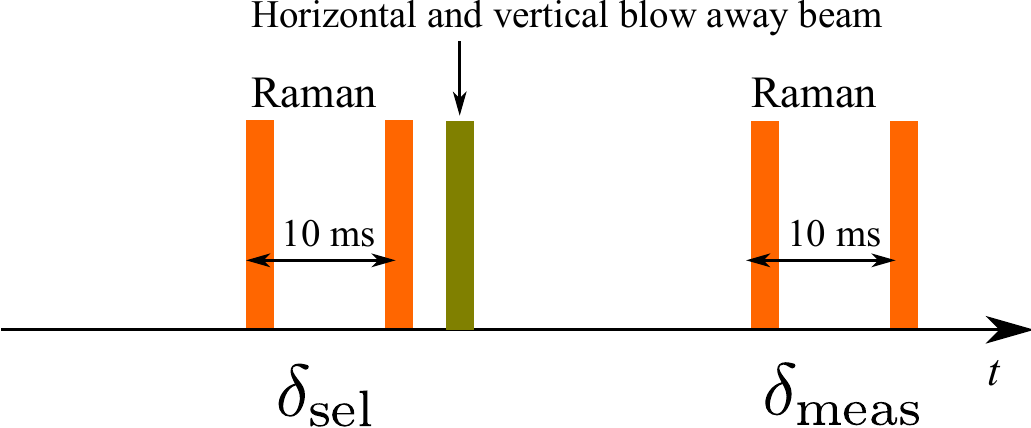}
\end{center}
\caption{\label{fig:sequence_interferometre_simple} Description of a velocity sensor based on a sequence of four $\pi/2$ pulses. The first two $\pi/2$ pulses, at the same frequency $\delta_\mathrm{sel}$, are separated by the Ramsey time $T_\mathrm{Ramsey}$. The last two ones,  at the frequency $\delta_\mathrm{meas}$, are separated by the same delay $T_\mathrm{Ramsey}$.}
\end{figure}

The idea of this atom interferometer is very similar to this scheme. Instead of using $\pi$ pulses, the Ramsey method of separated oscillatory fields is used and the $\pi$ pulse is replaced by two $\pi/2$ pulses. The sequence is described on  Fig.~\ref{fig:sequence_interferometre_simple}. 

After the first two $\pi/2$ pulses, the velocity distribution of atoms is now given by a Ramsey fringe pattern. The fringe spacing is inversely proportional to the Ramsey time $T_\mathrm{Ramsey}$ : $\Delta v = 2\pi/(2k T_\mathrm{Ramsey})$. The second set of $\pi/2$ pulses will then convolve this distribution with itself. The velocity of atoms will therefore be measured with a resolution that scales as $2\pi/(2k T_\mathrm{Ramsey})$ and the sensitivity is proportional to $2T_\mathrm{Ramsey}/\lambda$. The main difference with the previous approach is that the number of atoms is independent of the duration $T_\mathrm{Ramsey}$. It is therefore possible to increase the sensitivity, keeping the same signal-to-noise ratio.

\begin{figure}
\begin{center}
\includegraphics[width=.5\linewidth]{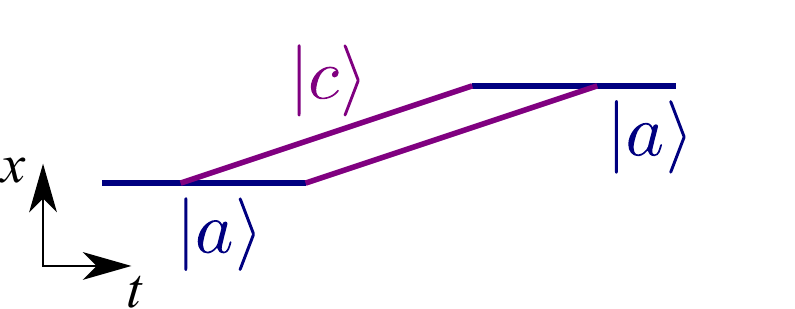}
\end{center}
\caption{\label{fig:diagramme_interferometre_Ramsey} Description of the semi-classical trajectories of atoms in the atom interferometer. Only the trajectories that play a role in the interferometer are displayed}
\end{figure}

This sequence of $\pi/2$ pulses realises an atom interferometer : each $\pi/2$ pulse acts as a beamsplitter and is used to separate or recombine atomic wave packets\cite{Borde1989}. The two semi-classical trajectories of atoms are depicted on Fig.~\ref{fig:diagramme_interferometre_Ramsey}. 

After the first two $\pi/2$ pulses, the distance between the two arms is $\Delta x = 2 v_r T_\mathrm{Ramsey}$. It is straight-forward to check that the sensitivity $1/\Delta v$ is obtained using the Heisenberg limits : $\Delta x \Delta v = h$. The larger the distance between the two arms of the interferometer, the larger the sensitivity. 

\subsection{Measurement scheme}
\label{sec:mes_scheme}
\begin{figure}
\begin{center}
\includegraphics[height=.3\linewidth]{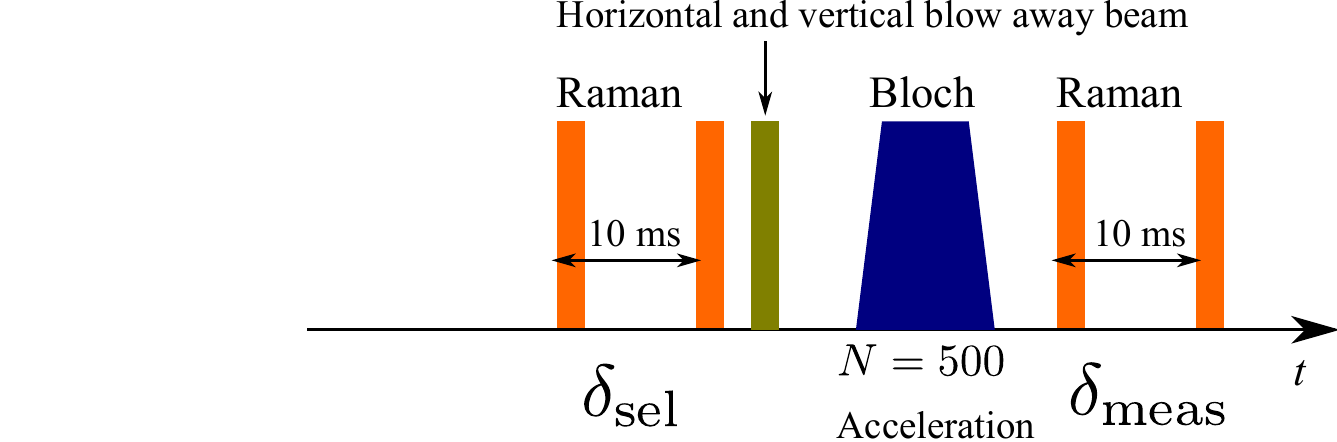}
\caption{\label{fig:sequencehsurmB} Description of the interferometer used to measure the atom recoil velocity. The Ramsey-Bord\'e interferometer is built out of four $\pi/2$ pulses (orange). In the middle of the interferometer, we use Bloch oscillations to transfer to atoms 1000 recoils(blue).}
\end{center}
\end{figure}

Figure~\ref{fig:sequencehsurmB} presents a simplified version of the temporal sequence of the pulses used in our atom interferometer. The Ramsey-Bord\'e interferometer is used to measure the velocity change induced by the Bloch oscillations. 
In a typical experiment, we transfer to the atoms 1000 recoils ($N=500$ Bloch oscillations). The Doppler shift induced by the Bloch oscillations is then : 

\begin{equation}
\delta = 2k_\mathrm{R}\times 2Nv_r = 4\hbar\frac{k_\mathrm{R}k_\mathrm{B}}{m}
\end{equation}
where $k_\mathrm{R}$ and $k_\mathrm{B}$ are the wave vector for the Raman and Bloch beam. For Rubidium atoms, this shift is about 15~MHz. 

We can calculate the relative sensitivity $\frac{v_r}{\Delta v_r}$ to the recoil velocity. It is given by the sensitivity in velocity and the number of transferred recoil
\begin{equation}
\frac{v_r}{\Delta v_r} = \frac{2Nv_r}{\Delta v}
\end{equation}
This sensitivity is also equal to the product of the Doppler effect and the Ramsey time. For $T_\mathrm{Ramsey} = 10\unite{ms}$, and $N=500$ Bloch oscillations, the sensitivity is about $1.5\times 10^5$. The sensitivity of the interferometer is similar to what is usually called the phase shift of the interferometer (in our case it will be $1.5\times 10^5\times 2\pi\unite{rad}$). But we should note that in our experiment (as in every atom interferometer), the total phase shift is close to 0 because the phase shift on the atoms is compensated by the frequency shift on the Raman laser beam. 

\subsection{The experimental setup}

\begin{figure}
\includegraphics[width=.9\linewidth]{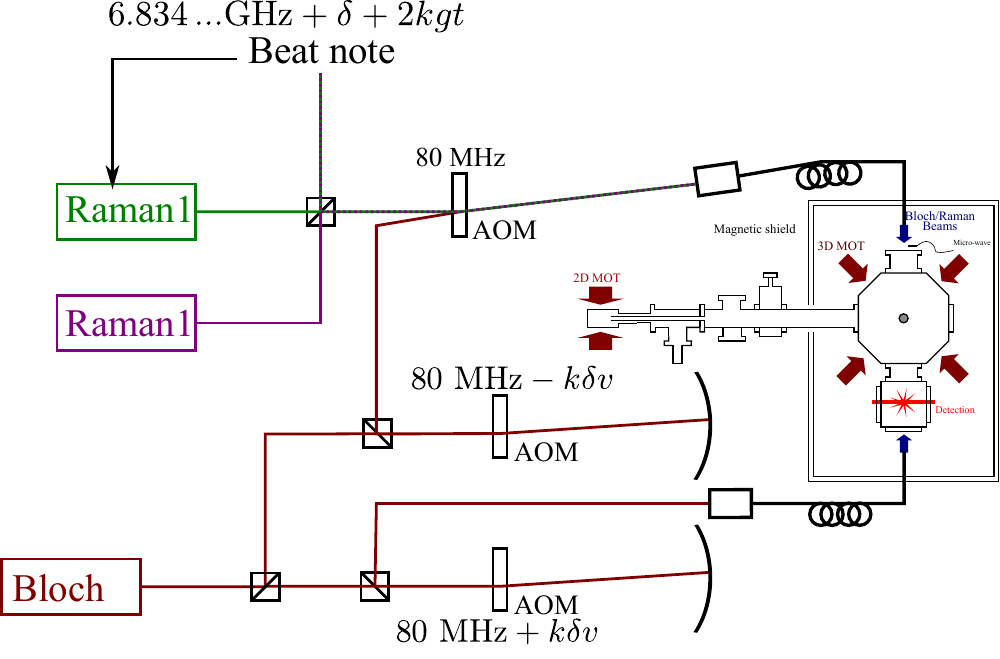}
\caption{\label{fig:schemaoptique} Schematic of the experiment : the laser system and the vacuum cell}
\end{figure}

Figure~\ref{fig:schemaoptique} shows a schematic view of the setup used to run the atom interferometer. In order to prepare a cold atomic cloud we use  a two-dimensional magneto-optical trap (2D-MOT) loading a 3D-MOT (during about 400~ms). The 3D-MOT is protected against stray magnetic fields by a magnetic shield. Atoms trapped in the 3D-MOT are further cooled using an optical molasses at a temperature close to 3~$\mu K$. After the atom interferometer, we measure the two populations of atoms in the two internal states by looking at their fluorescence in a detection zone which is placed about 20~cm below the trap. 

In order to perform the Raman transition, we use two phase-locked diode lasers (labeled Raman 1 and Raman 2 in  Fig.~\ref{fig:schemaoptique}). In order to precisely control the Raman frequency, we compare the beat notes of the two lasers to a well-controlled micro wave. The resulting beat frequency is then sent to a fix frequency phase-lock loop. In order to change the Raman frequency we simply change the micro-wave frequency. More specifically, we have to implement a continuous frequency sweep in order to compensate for the acceleration of gravity and a frequency jump to switch between $\delta_\mathrm{sel}$ and $\delta_\mathrm{meas}$.

The acceleration is a three step process. We first need to load the atoms into the lattice. Starting with atoms at rest with respect to the lattice, we adiabatically switch on the laser power. Then the lattice is accelerated and finally, with the lattice at a given speed, the laser power is adiabatically switched off. 

This sequence can be efficiently realized using acousto-optic modulators (AOM). An arbitrary waveform generator is used in order to control both the power and the frequency of the two AOMs. In order to change the velocity of the lattice by 2 recoil velocities ($12 \unite{mm/s}$), one needs to shift the frequency difference by about $30\unite{kHz}$ for Rb atoms. Because the frequency of each beam is changed and the AOMs are in a double-pass configuration, this corresponds to a change of frequency of about $7.5\unite{kHz}$. The typical bandwidth of an 80\unite{MHz} AOM in a double-pass configuration ($\pm 5\unite{MHz}$) allows to accelerate the lattice at a speed above 6\unite{m/s}, i.e. transferring more than 1000 recoil velocities. 


\subsection{Description of the experimental sequence}
The experimental sequence is slightly more complicated than the sequence described in Fig.~\ref{fig:sequencehsurmB}. In this scheme, the speed of the atoms at the end of the interferometer, after 500 Bloch oscillations, is several meters per second. This is not convenient for an efficient detection of atoms. In the experimental scheme, instead of starting the interferometer with atoms at small velocity to measure the recoil induce by 500 BOs, we start with atoms that are already accelerated and then transfer the 500 BOs in order to reduce their speed. The initial acceleration is done also with Bloch oscillations. By using the same number of BOs for the initial acceleration and the measurement, we end up with atoms having a velocity independent of the number of BOs. 

Furthermore, we use an atomic elevator in order to displace the center of the trajectory of the atoms. Using this method, we can run the interferometer close to the central part of the science chamber, where the systematic effects (mainly the magnetic field) are well controlled. 

\begin{figure}
\begin{center}
\includegraphics[width=.7\linewidth]{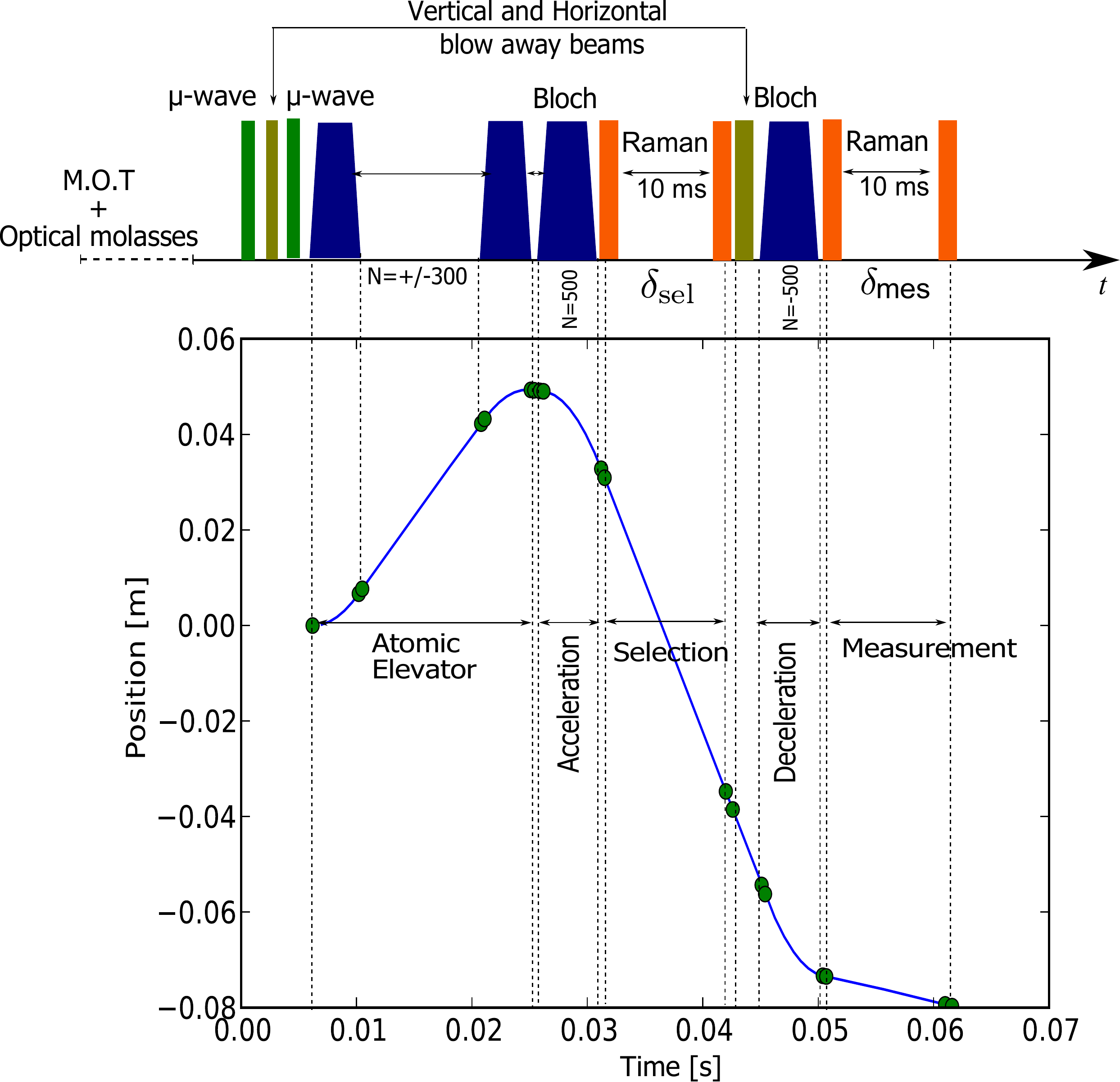}
\caption{\label{fig:sequencehsurmD} Description of the full sequence of the measurement of $h/m$. The Ramsey-Bord\'e interferometer is built out of four $\pi/2$ pulses (Orange). The Bloch oscillations used for the measurement, but also for the displacement of the cloud and the first acceleration, are in blue.}
\end{center}
\end{figure}

Figure~\ref{fig:sequencehsurmD} describes the full sequence used for the measurement. There is a total of four BO pulses (blue) and four $\pi/2$ pulses for the Ramsey-Bord\'e interferometer. Below the sequence, the trajectory of the atoms is depicted. This figure is a true scale description. The separation between the two trajectories (about 100\unite{\mu m}) cannot be seen compared to the displacement (about 5~cm) of the cloud.


The interferometer that we use is sensitive to the recoil velocity but also to any forces on atoms, especially the gravity. In order to remove the effect of gravity, we alternate two sequences with opposite directions of the acceleration. By taking the difference between the two measurements, we are able to cancel the uniform part of the gravity field. 

Furthermore, as explained in the first lecture, there is a light shift in the Raman transition. There is also a shift in the transition due to magnetic field. This shift of the resonance condition will induce an error in the velocity measurement. In the configuration of the interferometer that we use, the absolute frequency of the transition does not play a role because the phase of the interferometer only depends on the frequency shift of the Raman laser between the selection and the measurement (this is due to the fact that the time spent in each internal state is the same for both arms of the interferometer). However, non uniform shifts (both in time and space) will induce a systematic effect. In order to compensate the main part of this systematic, we exchange the direction of the Raman beams. Therefore, the relative sign between the Doppler effect and the systematic effect changes and we can cancel this effect. 

\begin{figure}
\begin{center}
\includegraphics[width=.49\linewidth]{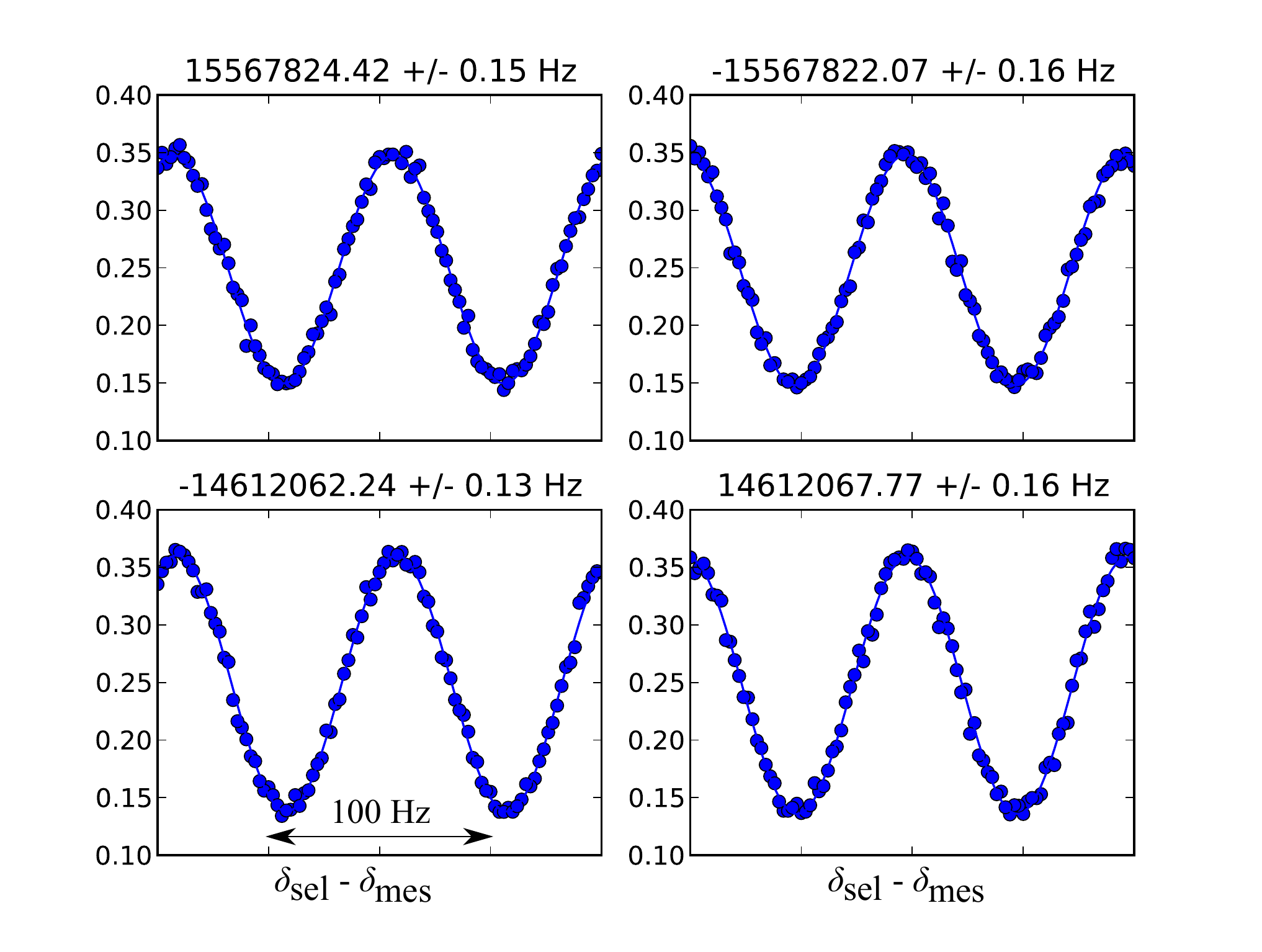}
\includegraphics[width=.49\linewidth]{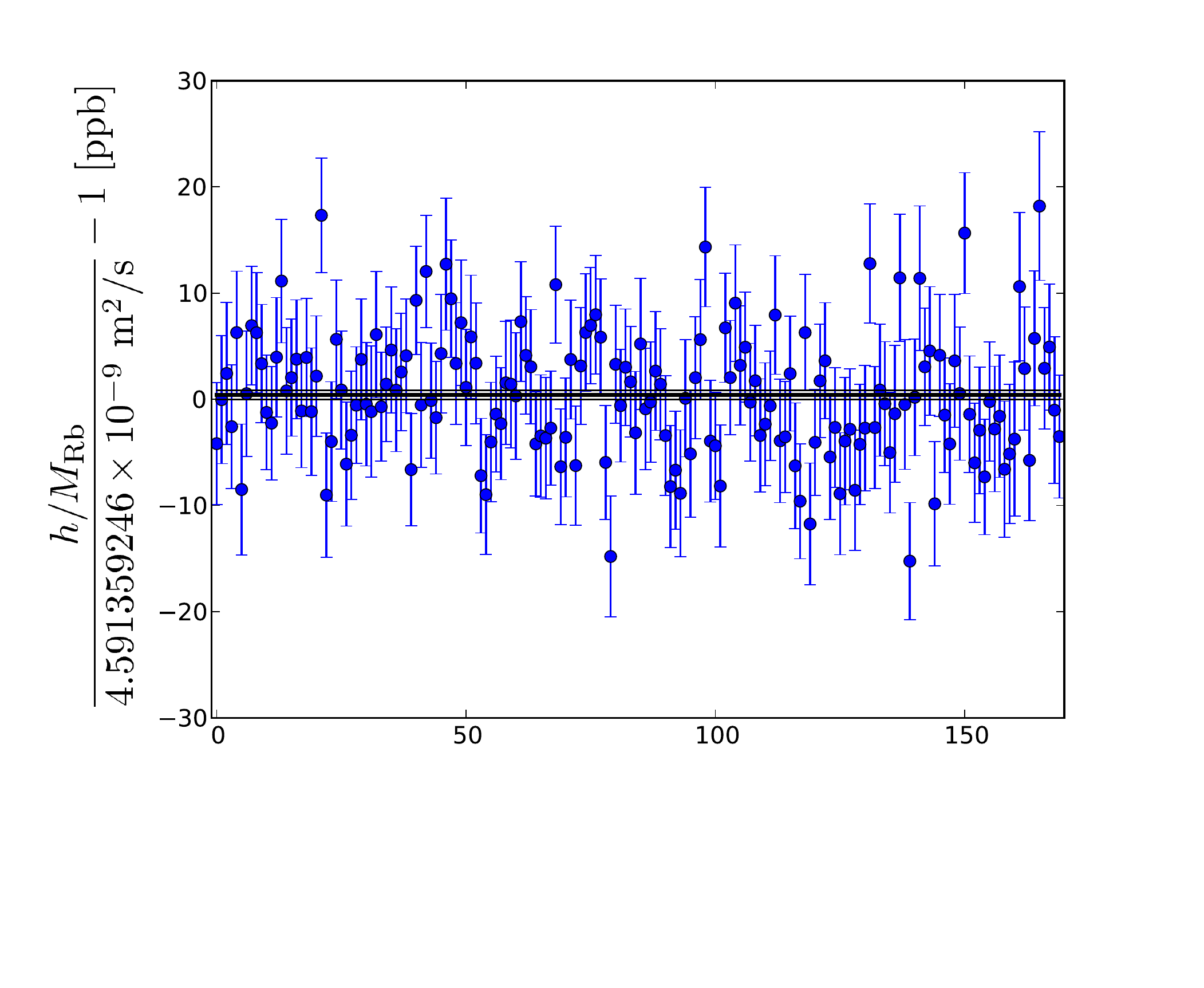}
\caption{\label{fig:typical_graph} Left : four spectra fitted with a sine function. Right : series of 150 measurements taken over 14 hours. Each measurement has a relative uncertainty of about $5\times10^{-9}$.}
\end{center}
\end{figure}

As a conclusion, for a measurement of the atomic recoil, we need to record four spectra (two Bloch directions times two Raman directions). Typical spectra are displayed on Fig.~\ref{fig:typical_graph}. As we can see, the Doppler shift is about 15~MHz and the fringe width 100~Hz. The precision at which we can fit the central fringe is about 0.15\unite{Hz} for 100 points. And by combining four spectra (400 points), we obtain a relative uncertainty of about $5\times10^{-9}$. It takes about 5 minutes to record this graph. 

\subsection{Systematics}
\label{sec:effet_sys}
\begin{table}
\caption{\label{BudgetError} Error budget on the determination of
$1/ \alpha$ (systematic effects and relative uncertainty in part per $10^{10}$). }
\begin{tabular}{lcc}
\multicolumn{1}{l}{Source}
&Correction  &\parbox{2cm}{Relative uncertainty }\\
\hline Laser frequencies& &1.3\\
Beams alignment&-3.3& 3.3\\
Wavefront curvature and Gouy phase&-25.1 & 3.0\\
2nd order Zeeman effect&4.0 & 3.0 \\
Gravity gradient&-2.0& $0.2$ \\
Light shift (one photon transition)& & 0.1\\
Light shift (two photon transition)& & 0.01 \\
Light shift (Bloch oscillation)& & 0.5 \\
Index of refraction atomic cloud & & \\
and atom interactions& &2.0 \\ \hline
Global systematic effects&-26.4 & 5.9\\ \hline 
Statistical uncertainty& & 2.0\\
Rydberg constant and mass ratio & &2.2 \\ \hline \hline
Total uncertainty& & 6.6\\
\end{tabular}
\end{table}

The main limitation of the experiments comes from systematic. We will briefly describe the systematics effects of the experiment. 

Table \ref{BudgetError} gives the error budget. The systematic effects are reduced compared to our previous measurements \cite{{Clade2},{Cadoret2008}}. The lasers are locked on a Fabry-Perot cavity stabilized with a standard laser and their frequencies are measured with a frequency comb to reduce the frequency uncertainties to less than 50 kHz. The maximum angle between the lasers used for the Raman transitions and the Bloch oscillations is estimated to 40 $\mu$rad from the coupling between the two optical fibers. Moreover this value has been confirmed by the observation of the effect of the misalignment between the Bloch beams. The effect of the Gouy phase and the wave front curvature, which has been reduced by increasing the waist of the laser beam from $w$ = 2 mm to $w$ = 3.6 mm, has been carefully controlled with a Shack-Hartmann wave front analyzer. The parasitic magnetic field has been reduced with a double magnetic shield and a precise mapping of the magnetic field gives now a relative correction of $4\times 10^{-10}$. Thanks to the good collimation of the laser beams, the section of the laser beams varies by about $4\times 10^{-3}$ along the atomic trajectory and the result is a very good cancellation of the light shift effects between the upward and downward trajectory. From the density of the cloud of cold atoms after the RF selection and the two first sequences of BO (about $2\times 10^8$ atoms/cm$^3$), the effects of the refractive index and of the interactions between the atoms are estimated at a $10^{-10}$ level, corresponding to a conservative uncertainty of $2\times 10^{-10}$ in Table \ref{BudgetError}.  Thanks to the double cell with a differential pumping, the effect of the refractive index due to the background vapor (about $10^7$ atoms/cm$^3$) is now at the negligible level of a few $10^{-11}$.

Taking into account all these corrections, the measured value of the ratio $h/m_{\mathrm{Rb}}$ is $4.591~359~2729~(57) \times 10^{-9}$m$^2$s$^{-1}$. 

In the next section, we will see how this value of $h/m_{\mathrm{Rb}}$ can be used to deduce a value of the fine structure constant $\alpha$, and contributes to precise test of QED.

\section{Determination of the fine structure constant using atom recoil measurement}

\subsection{The fine structure constant and quantum electrodynamics}

The fine structure constant, $\alpha$, was introduced in 1916 by A. Sommerfeld in order to explain the fine structure observed in hydrogen. A.~Sommerfeld tried to include the special relativity in the Bohr model\cite{Sommerfeld1916}. This dimensionless constant then represents the velocity $v_e$ of the electron in the first Bohr orbit divided by the speed of light. It is on the order of  $7\times 10^{-3}$.
\begin{equation}
\label{eq:alpha}
\alpha = \frac{v_e}{c} = \frac{e^2}{4\pi\epsilon_0 \hbar c}
\end{equation}

In 1916, A.~Sommerfeld did not succeed in precisely describing the fine structure of the hydrogen atom. The main reason comes from the spin of the electron that was not yet understood. The full explanation of the spin of the electron comes from the equation that P.M.~Dirac introduced in 1928 in order to write a relativistic version of the Schr\"odinger equation \cite{Dirac1928}. In this equation, Dirac introduced an internal degree of freedom, the spin angular momentum $\vect{S}$, which induces an intrinsic magnetic moment that is coupled to the electro-magnetic field.  This moment, $\vect{\mu_S}$, is proportional to the Bohr magneton $\mu_B = \frac e{2m_e}$ and the spin $\vect{S}$:
\begin{equation}
\vect{\mu_S} = -g_e\mu_B\frac{\vect{S}}\hbar
\label{eq:lande_factor}
\end{equation}
where $g_e$ is the dimension-less Land\'e g-factor. The Dirac equation was able to predict that the g-factor of the electron was two : $g_e=2$. 

\begin{figure}
\begin{center}
\includegraphics[width=.5\linewidth]{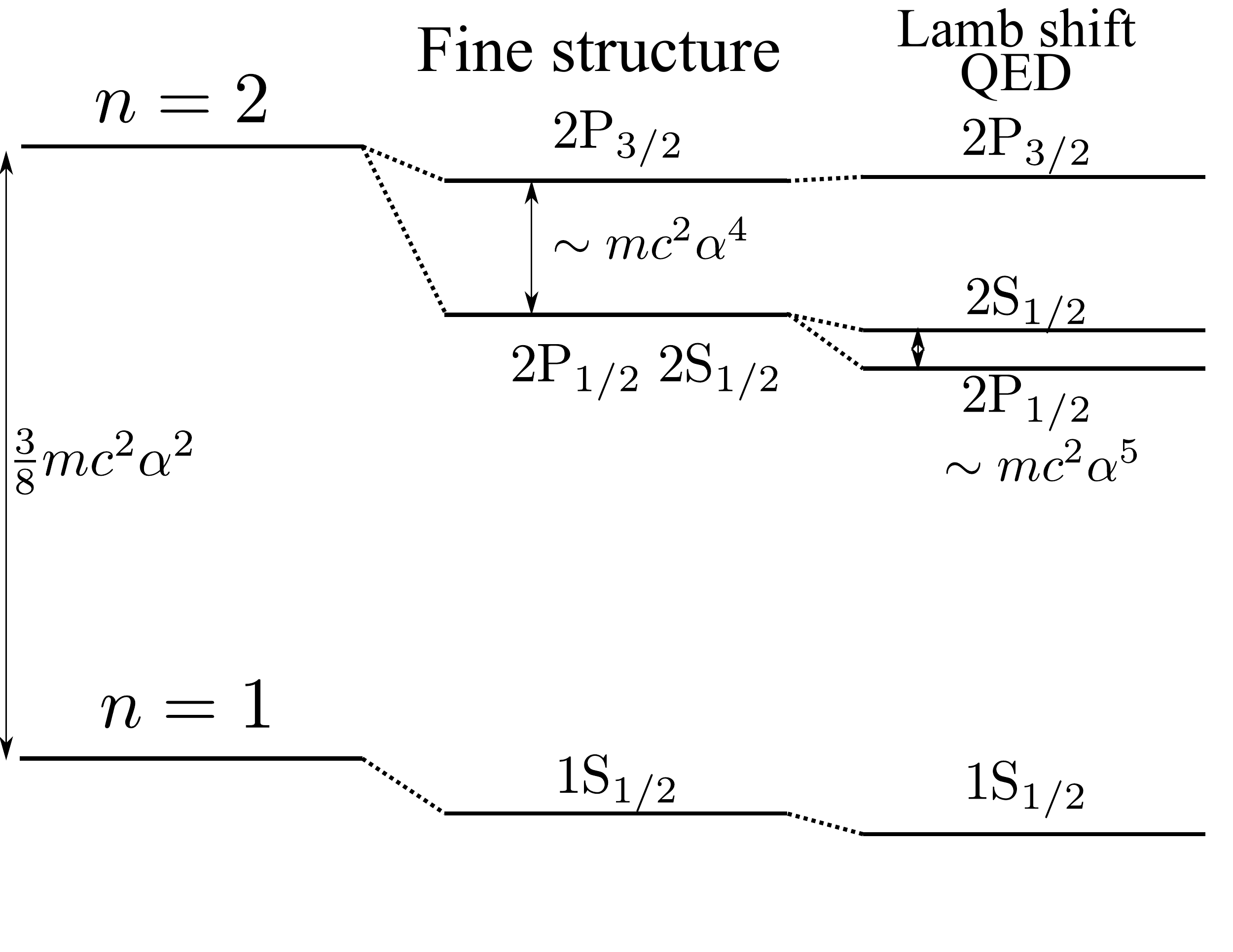}
\end{center}
\caption{\label{fig:hydro}Structure of the hydrogen atom. The energy levels are set by the Rydberg constant ($R_\infty = \frac12mc^2\alpha^2$), where $\alpha$ is the fine structure constant.}
\end{figure}

This equation also succeed in the calculation of the fine structure of the hydrogen (see Fig.~\ref{fig:hydro}). The scaling factor was the same fine structure constant that Sommerfeld introduced earlier. 

The atomic structure of hydrogen and the calculation of the Land\'e factor of the electron are two of thegreatest achievement of the Dirac equation. However in the late 1940's, this theory was superseded with the quantum electrodynamics theory (QED). The first evidence of an effect beyond the Dirac equation is due to Lamb and Retherford\cite{Lamb1947}, they discovered that the 2S and 2P states of hydrogen atoms are not degenerate (see Fig.~\ref{fig:hydro}). This effect is now called the Lamb shift. The second effect concerns the magnetic moment of the electron. Using precise measurement of the hyperfine structure of alkaline atoms, Kusch and Foley \cite{Kusch1948} discovered that the Land\'e factor of the electron was slightly higher than 2. The anomalous magnetic moment of the electron, $a_e = (g_e-2)/2$ was then introduced. 

Simultaneously, theoretical work was done in order to have a full quantum description of the interaction between light and matter. The first formulation was due to Dirac\cite{Dirac1927}, however this theory relies only on the first order of perturbation theory. At higher orders in the series infinities emerged, making such computations meaningless. In 1947, Bethe was able to remove those infinities using the so called renormalisation theory. He was then able to calculate the first order correction in the Lamb shift of hydrogen. Following Bethe's work, Schwinger was able to calculate the first order correction in the Land\'e factor of the electron\cite{Schwinger1948}. 

Over the 60 years that separate the birth of quantum electrodynamics and the present days, the theory has been used to describe phenomena ranging from atomic physics to astro-physics. It has been precisely verified in many systems. The two first QED effet that were observed (the Lamb shift and the anomalous magnetic moment of the electron) are still nowadays the most precisely measured QED effect. They make the QED the most precisely tested physical theory.  
We will now focus on the anomalous magnetic moment of the electron. 

\subsection{The anomalous magnetic moment of the electron}


In his paper of 1948\cite{Schwinger1948}, Schwinger was able to calculate the first order QED correction to the g-factor of the electron. He found that:
\begin{equation}
\frac{g_e}2 = 1 +\left(\frac\alpha{2\pi}\right)
\end{equation}
This equation can be extended to any order in $\alpha$:
\begin{equation} 
\label{eq:dev_g_2}
\frac {g_\mathrm{e}}2 = 1+C_1\left(\frac\alpha\pi\right) +
C_2\left(\frac\alpha\pi\right)^2 + C_3\left(\frac\alpha\pi\right)^3 +
C_4\left(\frac\alpha\pi\right)^4 + ...
\end{equation}

\begin{figure}
\begin{center}
\includegraphics[width=.49\linewidth]{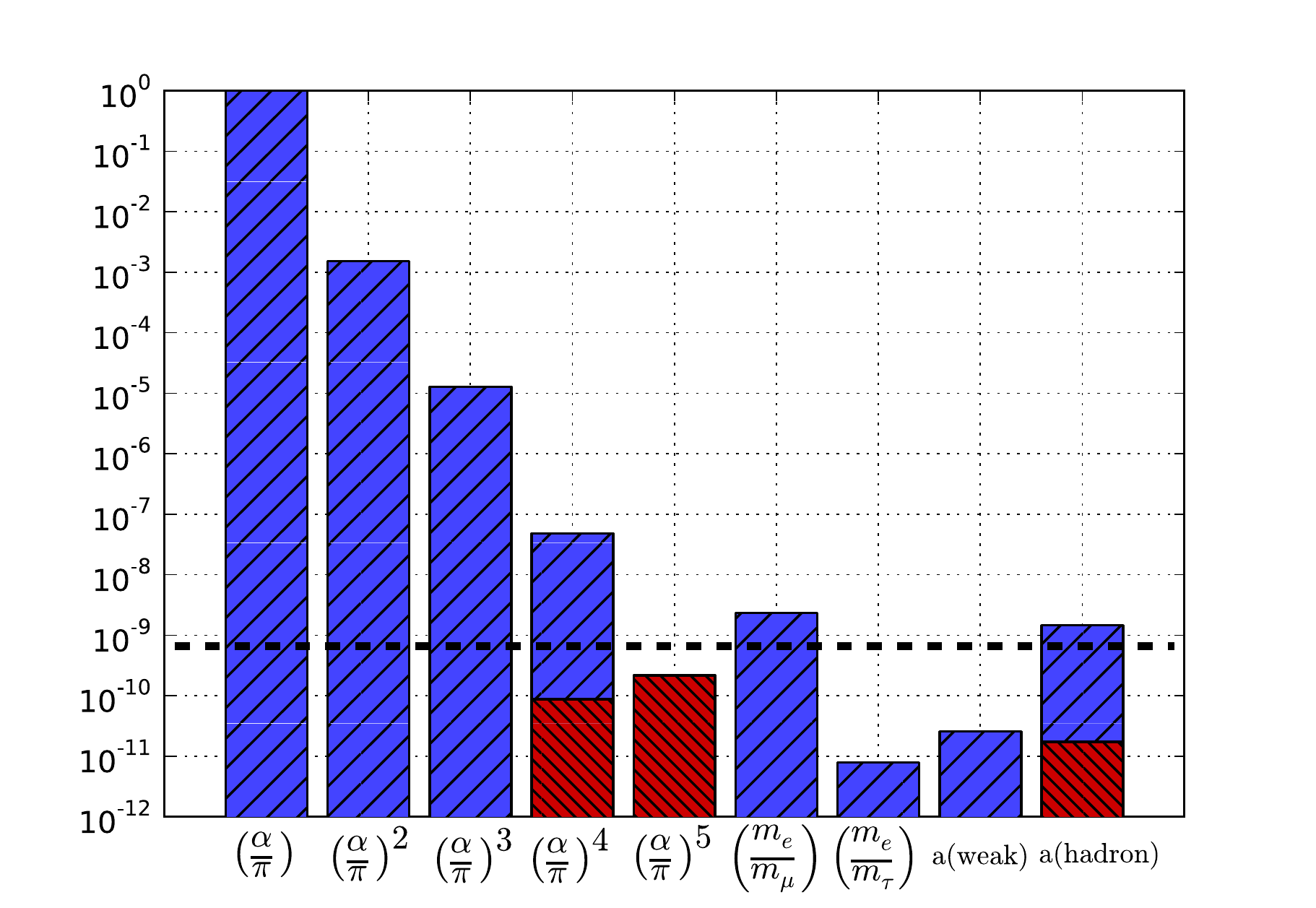} \hfill \includegraphics[width=.49\linewidth]{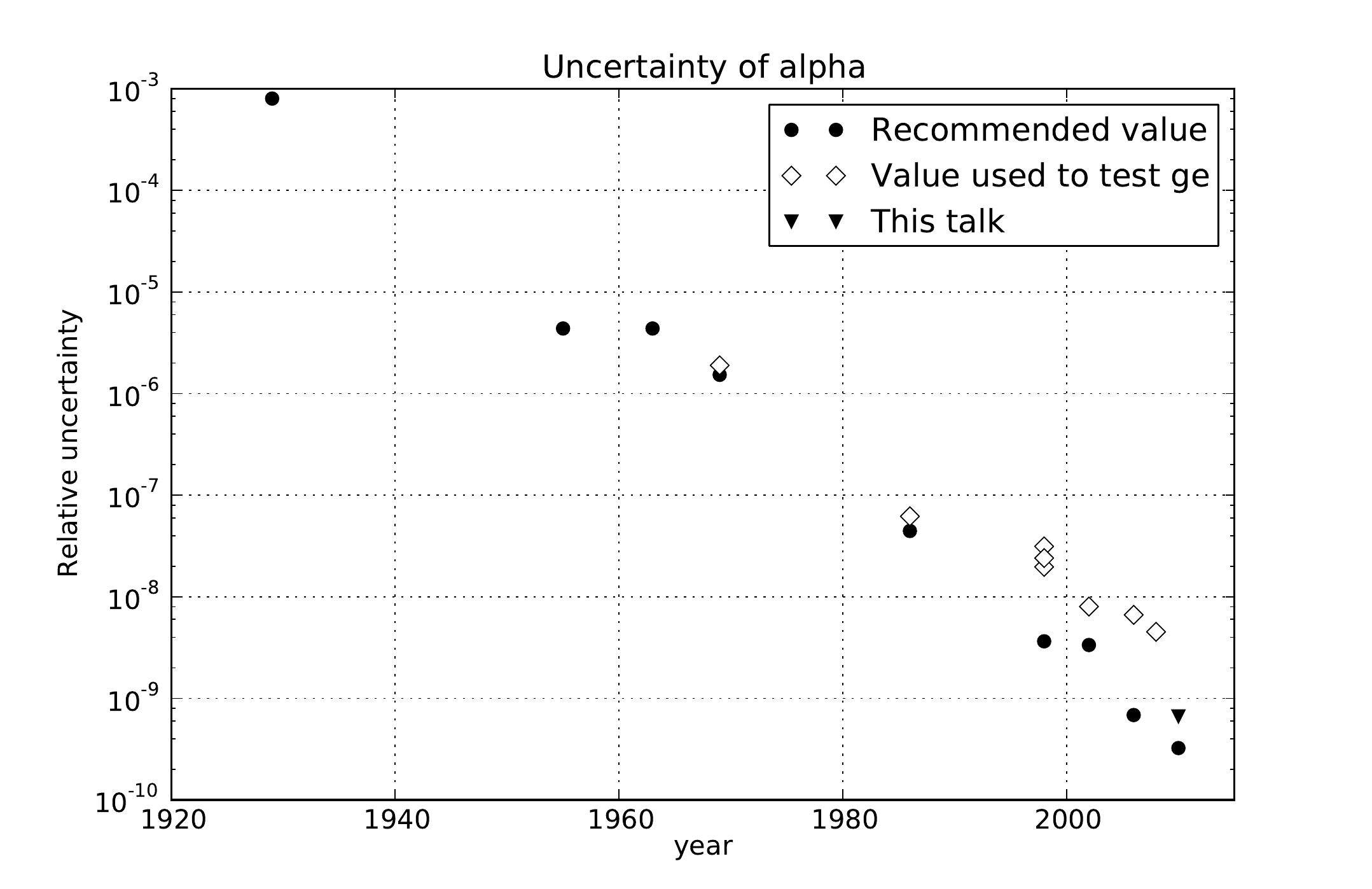}
\end{center}
\caption{Left : Contributions to the g-factor of the electrons. In blue, the amplitude of the contribution, in red its associated uncertainty. Right : history of the measurement of the g-factor of the electron and the best value of $\alpha$ used to test QED.}
\label{fig:cont_ae}
\end{figure}

The calculation of the coefficients becomes increasingly complex with the order in $\alpha$. A complete history of the calculation of those coefficients can be found in Ref.~\cite{Kinoshita2010}. In 2012, the $C_5$ coefficient was computed by the group of T.~Kinoshita\cite{Aoyama2012}. Using an automatic code generator, they evaluated the 12\,672 diagrams of the tenth-order diagrams\cite{Aoyama2012}.  Figure~\ref{fig:cont_ae} shows the amplitude of the different known contributions to the g-factor of the electron. The first three main contributions are known analytically, while the fourth and fifth are only calculated numerically. Those computed terms have an uncertainty, on the order of a few parts in $10^{-11}$  and which is depicted in red. In eq.~\ref{eq:dev_g_2}, we have only written the electronic QED contribution. Other contributions are present, at the order of $10^{-9}$ and below. The lepton contributions (due to muons and taus) and also the electro-weak and hadronic contribution. They are shown on Fig.~\ref{fig:cont_ae}, with their uncertainties (red). 

The Land\'e factor of the electron has been introduced in eq.~\ref{eq:lande_factor}, by comparing its magnetic moment to the Bohr magneton. A direct way of measuring this factor is done by comparing the cyclotron frequency and the Larmor precession frequency of the electron in a magnetic field $B$:
\begin{eqnarray}
\omega_\mathrm{cyc} &=& \frac{eB}m \\
\omega_\mathrm{lar} &=& \mu_S B
\end{eqnarray}

The Land\'e factor of the electron is now measured at the level of $10^{-12}$. Progresses in this measurement are due to the work of Dehmelt \cite{VanDick} who was able to trap a single electron and then to the work of Gabrielse \cite{Hanneke2011} in which the electron is cooled down to cryogenic temperatures and is in the ground state of the Penning trap. The Land\'e factor of the electron is measured with an uncertainty of $0.28\times 10^{-12}$. This uncertainty corresponds to an uncertainty on $\alpha$ which is equal to $0.24\times 10^{-9}$.

\subsection{Determination of $\alpha$ using the recoil velocity}

The Rydberg constant can be written as a function of the fine structure constant $\alpha$ using the formula:
\begin{equation}
R_\infty = \frac{m_e}{2hc} c^2\alpha^2.
\label{eq:rydberg}
\end{equation}
The Rydberg constant is known with an uncertainty of $5 \times 10^{-12}$. In order to get a value of $\alpha$, we need to know the ratio $m_e/h$. Because we measure $h/m$, we need the ratio $m/m_e$ in order to get the ratio $m_e/h$. This ratio is determined through the mass $m_p$ of a proton using the ratios $m/m_p$ and $m_p/m_e$, known  with uncertainties of $1.4\times10^{-10}$ (for rubidium) and $4\times10^{-10}$ respectively\cite{CODATA12}.

From the measurement of $h/m_\mathrm{Rb}$ presented in the previous section and using the CODATA value for the Rydberg constant and mass ratios, we obtain a value of $\alpha$ from which a value of the anomalous magnetic moment of the electron can be deduced :  

\begin{eqnarray}
\alpha^{-1}[\mathrm{LKB-2010}] &=& 137.035\,999\,037\,(91)\ \ [6.6\times 10^{-10}] \\
a_\mathrm{e}[\mathrm{LKB-2010}] &=& 0.001\,159\,652\,181\,88\,(78)
\end{eqnarray}


\begin{figure}
\includegraphics[width=.49\linewidth]{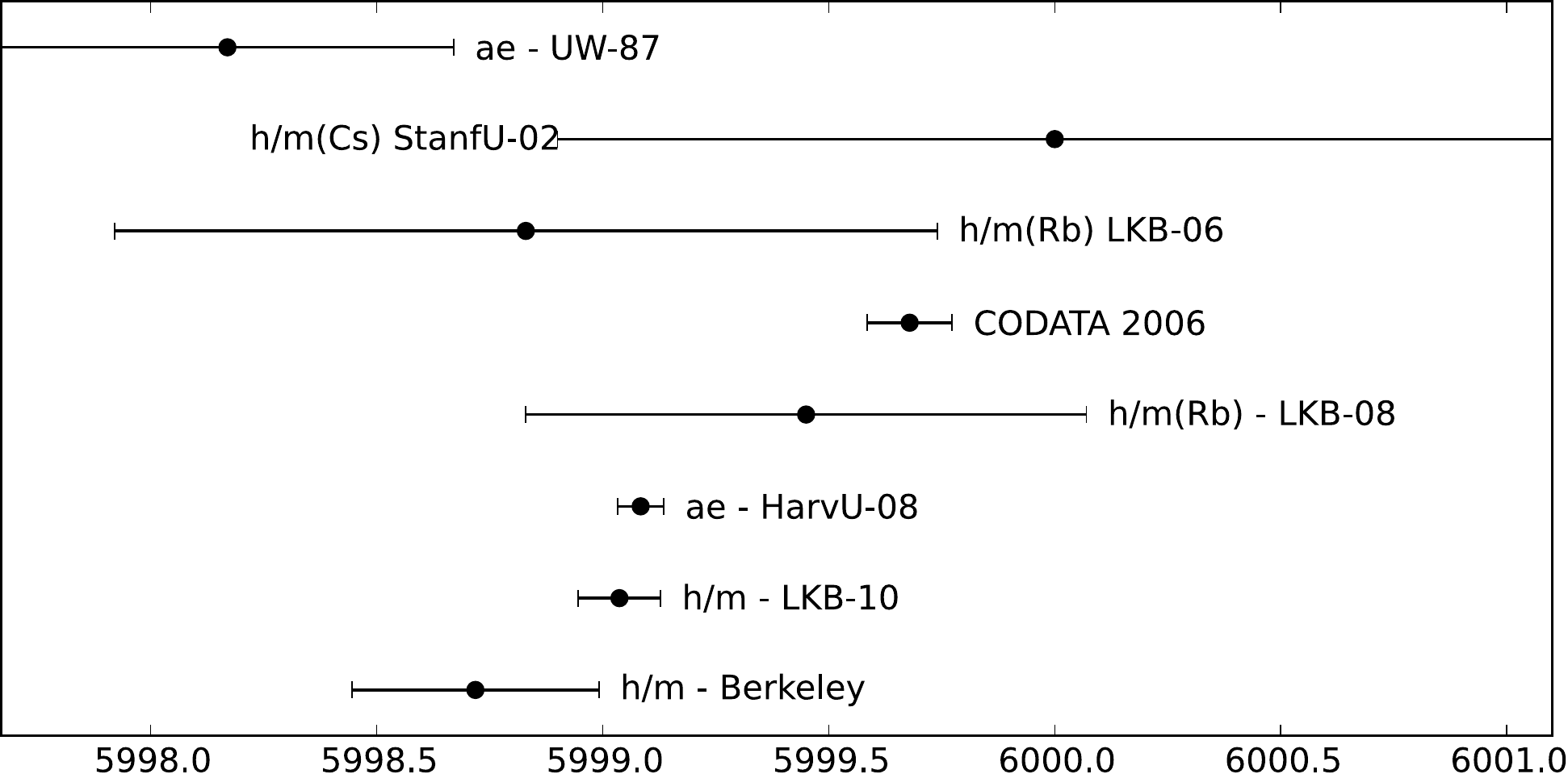}
\includegraphics[width=.49\linewidth]{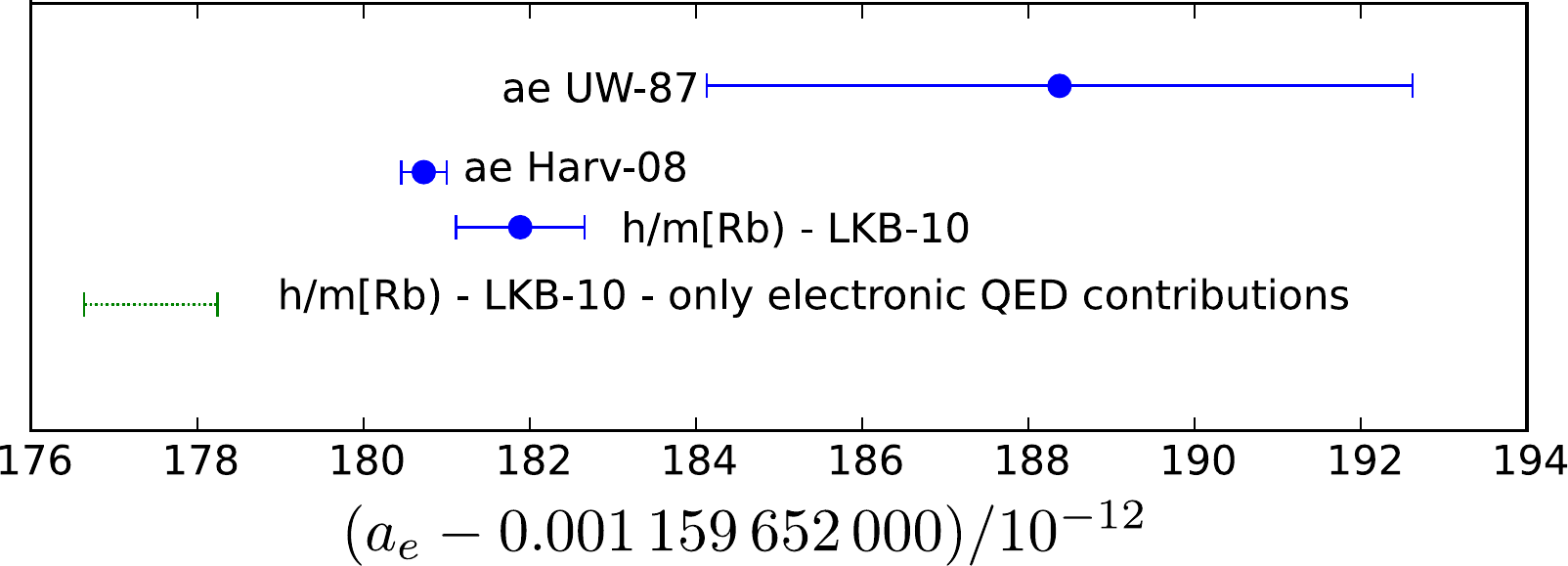}
\caption{\label{fig:comp} Comparison between the measurement of the magnetic moment of the electron and the recoil velocity measurement. Labels : UW-87, University of Washigton, group of Dehmelt \cite{VanDick}; StanfU-02, Stanford University, group of S.~Chu \cite{Wicht:02}; LKB(06, 08, 10), Laboratoire Kastler Brossel, group of Biraben \cite{clade:033001, Cadoret2008, Bouchendira2011}; CODATA 2006 \cite{CODATA2006}; HarvU-08, Harvard University, group of G.~Gabrielse \cite{Hanneke2011}; Berkeley 13, group of H.~M\"uller \cite{Lan2013}. 
Left : comparison using the fine structure constant. Right : comparison using the magnetic moment of the electron. We have included a value of $a_e$ deduced from the fine structure constant and equation \ref{eq:dev_g_2}, by taking into account only the electronic QED (i.e. without taking into account the muonic and hadronic contribution).}
\end{figure}

The upper part of Fig.~\ref{fig:comp} shows the comparison between the current values of $a_e$. The accuracy of the value of
$\alpha$ from the measurement of $h/m_\mathrm{Rb}$ is sufficient to test the contributions due to the muons and hadrons in the theoretical value of $a_e$. It is interesting to remark that in this experiment, it is possible to see the influence of high energy physics using low energy experiments. The comparison of the experimental value of $a_e$ with the calculated one using our measurement of $\alpha$ is the most precise test of QED, as we have reached the tenth order term. 

It is so accurate that one can
think, in a near future, of using these lab-size experiments
to check theoretical predictions tested up to now only on
particle accelerators (for example the existence of an internal
structure of the electron\cite{Gabrielse2006}).

One should also note that another precise test of QED is obtained through the measurement of the Lamb shift in the hydrogen atom. The Lamb shift (the energy splitting between the $^2S_{1/2}$ and $^2P_{1/2}$ states) is determined by the comparison of measurements of optical transitions in hydrogen. The Lamb shift is mainly a QED effect. However it is also affected by the finite size of the proton. A recent experiment measuring the proton radius using muonic hydrogen yielded to a value that is significantly different from the value that fits the QED calculations and the measurements in regular hydrogen \cite{Pohl2010}. The very precise test of QED done with the anomalous magnetic moment of the electron puts constraints of models that could explain the muonic hydrogen experiment. It remains therefore important to strengthen this test of QED with an improved QED-independent determination of $\alpha$.

\subsection{The ratio $h/m$ and the redefinition of the SI units}

In the redefinition of the SI planned by the CGPM in 2015,
the definition of the second will stay the same and the
kilogram will be defined by fixing the value of the Planck
constant $h$. This definition will be based on fundamental
constants and therefore the resolution of the CGPM
explicitly relies on the CODATA for the new definition \cite{GCWM2011}. 

The main challenge for the redefinition of the Kilogram,
and the main reason why this redefinition has been
delayed for several years, is the lack of a reliable link between
the microscopic and macroscopic masses. This link
is established with a relative uncertainty of $3\times10^{-8}$ and with large discrepancies between the different methods
(watt balances \cite{Steiner2005} and Avogadro project\cite{Andreas2011}). One can notice that the recently measured value of the Avogadro constant, which is the most accurate input datum for the kilogram
redefinition, is midway between the watt-balance
values \cite{Becker2012}.

\begin{figure}
\begin{center}
\includegraphics[width=.5\linewidth]{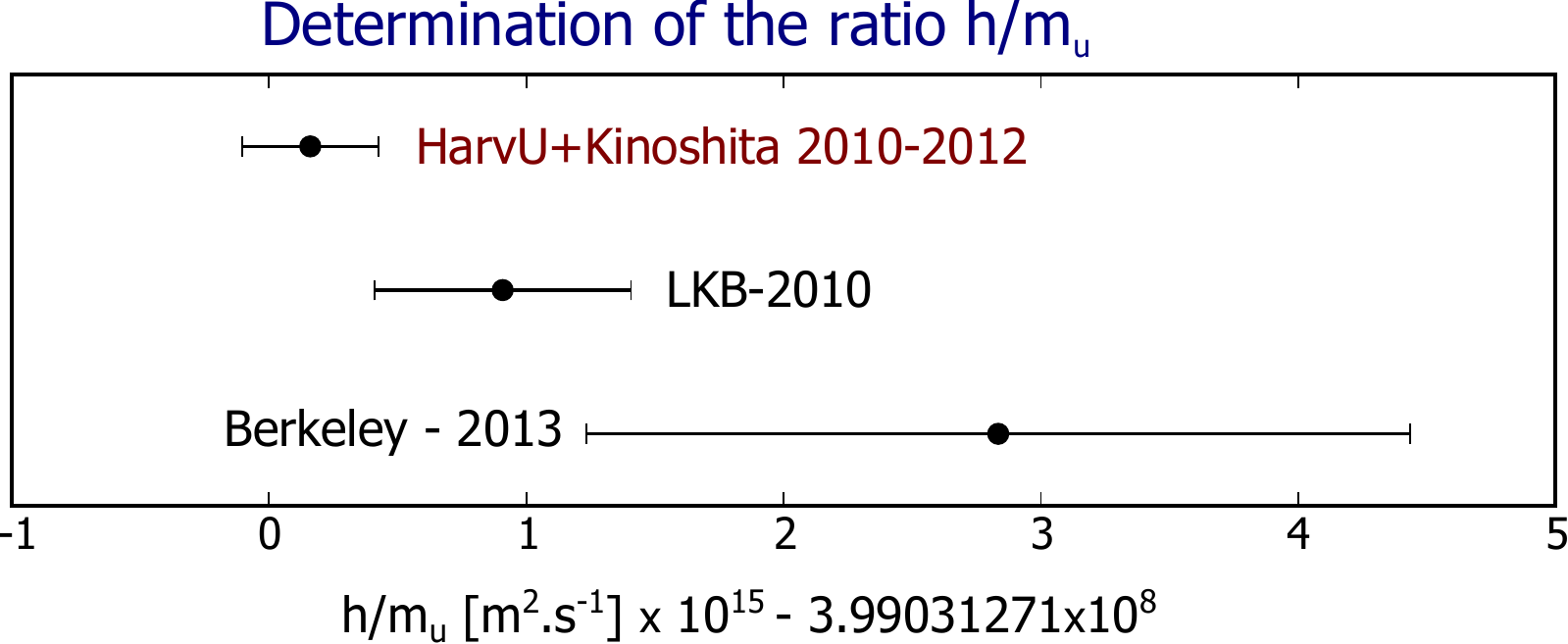}
\end{center}
\caption{\label{fig:comp_h_mu} Determinations
of the ratio $h/m_u$ deduced from the measurement of the rubidium
recoil \cite{Bouchendira2011} and Compton frequency of the cesium atom\cite{Lan2013}. The
most precise determination comes from the value of the fine
structure constant given by the experimental value of the $a_e$
measurement \cite{Hanneke2011} and QED calculations \cite{Aoyama2012}.}
\end{figure}

The watt balances measure the value of the Planck constant $h$. The Avogadro project measures the Avogado number $N_A$. They can be compared because the product $hN_A$, known as the Avogadro Planck constant is known precisely. Indeed we have :
\begin{equation}
\label{eq:molplanck}
hN_A = \frac{h}{m_u}\frac{M\left(^{12}C\right)}{12}
\end{equation} where $M\left(^{12}C\right) =12\times 10^{-3} \unite{kg/mol}$ is the carbon molar mass and $N_A$ is the Avogadro constant. The product $h N_A$ is, in
the current SI, equivalent to the ratio $h/m_u$. It seems more relevant to
us to consider $h/m_u$ in the framework of
the redefinition of the kilogram. In the future SI units,
the Avogadro constant $N_A$, which is used by chemists
to quantify and identify an amount of substances with
atoms and molecules, will be fixed. This will break the link
between atomic masses and molar masses. Consequently
$M(^{12}C)$ will no longer be equal to 12 g/mol, but will be
determined from equation \ref{eq:molplanck} using the ratio $h/m_u$. We have plotted on Fig.~\ref{fig:comp_h_mu} the three most precise determination of $h/m_u$ (coming from atomic measurements of $h/m$ or the experimental value of the $a_e$
measurement with QED calculations). The three values have an uncertainty below $4\times 10^{-9}$. These uncertainties are negligible compared to the uncertainties (about $3\times 10^{-8}$) involved in the redefinition of the kilogram and the ratio $h/m_u$ can be considered well known. 

The precisions of the determinations of $\alpha$ or $h/m_u$ is not a limiting factor for the redefinition. However, in the proposed new International Systems of Units, many  physical constants will have a fixed value. The constant $\alpha$ will be a keystone of the proposed SI, as many of the remaining
constants will strongly depend on its knowledge (such as
the vacuum permeability $\mu_0$, the von Klitzing constant $R_K$,
...) \cite{Mills2011}. For example, the same way that the weight of one liter of water is not exactly one kilogram, the value of $\mu_0$ will no longer be exactly $4\pi \times 10^{-7}$ in the new SI units. It will be a measured constant which will only depends on the fine structure constant. The value of $\mu_0$ in the new SI will therefore differ from $4\pi \times 10^{-7}$ by the difference between the value of $\alpha$ and its determination used for the redefinition. The same argument also applies to the ratio $h/m_u$ and the molar mass of $^{12}C$ which will no longer be $12\unite{g/mol}$ in the new SI.

Improving the measurement of $\alpha$ or $h/m_u$ will therefore improve the numerical value that will be fixed in the new SI. After the redefinition, those measurements will improve the knowledge of new fundamental constants. Especially, the measurement of $h/m_u$ will be used to directly link the atomic mass units to the SI units. 


\bigskip

\acknowledgments

The course presented at this Enrico Fermi school is the result of the research effort undertaken in the group of F.~Biraben at Laboratoire Kastler Brossel since 1998. The author of this paper thanks F.~Biraben and all the other contributors to this research (in alphabetic order): M.~Andia, R.~Battesti, R.~Bouchendira, M.~Cadoret, S.~Guellati-Khélifa, R.~Jannin, L.~Julien, E.~de.~Mirandes, F.~Nez, C.~Schwob. 

\bibliographystyle{varenna}
\bibliography{Pierre_sans_url}

\end{document}